# Roto-flexoelectric coupling impact on the phase diagrams and pyroelectricity of thin SrTiO$_3$ films

A.N. Morozovska[1,2], E.A. Eliseev[3], S.L. Bravina[1], A.Y. Borisevich[4], and S.V. Kalinin[5]

[1] Institute of Physics, National Academy of Science of Ukraine,
46, pr. Nauki, 03028 Kiev, Ukraine

[2] Institute of Semiconductor Physics, National Academy of Science of Ukraine,
45, pr. Nauki, 03028 Kiev, Ukraine

[3] Institute for Problems of Materials Science, National Academy of Science of Ukraine,
3, Krjijanovskogo, 03142 Kiev, Ukraine

[4] Materials Sciences and Technology Division, Oak Ridge National Laboratory,
Oak Ridge, TN 37831, USA

[5] Center for Nanophase Materials Science, Oak Ridge National Laboratory,
Oak Ridge, TN, 37831, USA

**Abstract**

The influence of the flexoelectric and rotostriction coupling on the phase diagrams of ferroelastic-quantum paraelectric SrTiO$_3$ films was studied using Landau-Ginzburg-Devonshire (LGD) theory. The phase diagrams in coordinates temperature - film thickness were calculated for different epitaxial misfit strains. Tensile misfit strains stimulate appearance of the spontaneous out-of-plane structural order parameter (displacement vector of an appropriate oxygen atom from its cubic position) in the structural phase. Compressive misfit strains stimulate appearance of the spontaneous in-plane structural order parameter. Gradients of the structural order parameter components, which inevitably exist in the vicinity of film surfaces due to the termination and symmetry breaking, induce improper polarization and pyroelectric response via the flexoelectric and rotostriction coupling mechanism. Flexoelectric and rotostriction coupling results in the roto-flexoelectric field that is antisymmetric inside the film, small in the central part of the film, where the gradients of the structural parameter are small, and maximal near the surfaces, where the gradients of the structural parameter are highest. The field induces improper polarization and pyroelectric response. Penetration depths of the improper phases (both polar and structural) can reach several nm from the film surfaces. An improper pyroelectric response of thin films is high enough to be registered with planar-type electrode configurations by conventional pyroelectric methods.

**Introduction**

Thin ferroic films, their multilayers and heterostructures attract permanent fundamental scientific interest and contain the great potential for multiple conventional and novel applications in nanoelectronics, sensorics and optics [1, 2, 3]. Unique functionalities of thin films are believed to originate primary from the surfaces and interfaces influence on the physical properties via great versatility of the coupling mechanisms enhanced or induced by the gradient effects and symmetry breaking [4, 5]. Since purely empirical way of the film properties study is extremely time-consuming to explore fundamental mechanisms and novel applications, theoretical studies of the surface and interface induced couplings are necessary for effective search of thin films with improved functional properties.

Elastic field gradients can induce polarization near the surfaces and interfaces via the flexoelectric coupling [6, 7, 8]. Note, that all materials are flexoelectrics [9], and all perovskite oxides with static rotations (such as octahedral rotations also called ***antiferrodistortions***) possess rotostriction and rotosymmetry [10]. The joint action of flexoelectric effect and rotostriction (previously named roto-flexo-effect) can lead to spontaneous polarization at an interface across which the octahedral rotations changes [11, 12]. In particular, it was predicted that the roto-flexo field at oxide interfaces gives rise to improper ferroelectricity and pyroelectricity in the vicinity of antiferrodistortive boundaries and elastic twins in bulk $SrTiO_3$ [11] as well as in the vicinity of semi-infinite $SrTiO_3$ surface below 105 K [12].

In the work we theoretically study the influence of the roto-flexoelectric coupling on the phase diagrams of thin $SrTiO_3$ films. Several mechanisms of the surface influence on the phase transitions in ferroelastic-quantum paraelectric films were included in our calculations, namely misfit strain between the epitaxial film and substrate, surface energy, rotostriction and electrostriction effects, and flexoelectric coupling. All the mechanics can essentially influence on the phase diagrams of thin $SrTiO_3$ films. It is well-known that misfit strain can induce ferroelectric polarization in quantum paraelectric $SrTiO_3$ [13, 14, 15, 16]. Within LGD phenomenology the surface energy coefficients determine so-called extrapolation lengths [17, 18]. Extrapolation lengths in dependence on their sign and value, which in turn are interface- and termination-dependent [19, 20], can enhance or suppress polarization via the spatial confinement of the film interfaces. Rotostriction and electrostriction effects lead to renormalization (in particular enhancement) of the biquadratic coupling between the structural order parameter (displacement vector of an appropriate oxygen atom from its cubic position in the oxygen octahedra) [13]. We demonstrated that the flexoelectric and rotostriction coupling, which are the strongest in the gradient regions of thickness

~5 nm under the film surfaces, induces improper polarization and pyroelectric response of thin SrTiO$_3$ films.

## 2. Problem statement and basic equations

Here we consider a thin epitaxial ferroelastic-quantum paraelectric film deposited on a rigid substrate. The film structure with capacitor geometry shown in **Fig.1a** corresponds to the short-circuited electrical boundary conditions. The structure shown in **Fig.1b** corresponds to the open-circuited electrical boundary conditions. The misfit strain $u_{11} = u_{22}\big|_{x_3=0} = u_m$ originates from the lattice mismatch between the epitaxial film and substrate.

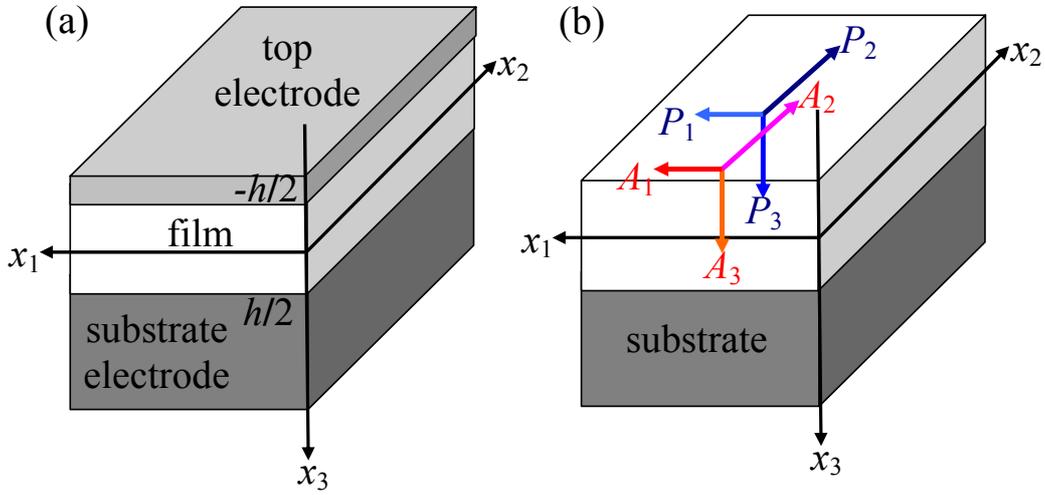

**Figure 1.** Ferroelastic film clamped on a rigid substrate. The film structure (a) corresponds to the short-circuited electrical boundary conditions; (b) corresponds to the open-circuited electrical boundary conditions.

In the parent high temperature phase above the structural phase transition the spontaneous order parameters are absent and the free energy has the form:

$$F = \int_V d^3r f_b(\mathbf{r}) + \int_S d^3r f_S(\mathbf{r}) \tag{1a}$$

The bulk energy density has the form [11]:

$$f_b = \begin{pmatrix} \beta_i(T)P_i^2 + \beta_{ij}P_i^2 P_j^2 + ... + \dfrac{1}{2}g_{ijkl}\left(\dfrac{\partial P_i}{\partial x_j}\dfrac{\partial P_k}{\partial x_l}\right) - P_i E_i - \dfrac{\varepsilon_0 \varepsilon_b \mathbf{E}^2}{2} \\ -q_{ijkl}u_{ij}P_k P_l + \dfrac{c_{ijkl}}{2}u_{ij}u_{kl} + \dfrac{1}{2}f_{ijkl}\left(u_{ij}\dfrac{\partial P_k}{\partial x_l} - P_k\dfrac{\partial u_{ij}}{\partial x_l}\right) + \xi_{ijkl}P_i P_j A_k A_l \\ + \alpha_i(T)A_i^2 + \alpha_{ij}A_i^2 A_j^2 + \dfrac{v_{ijkl}}{2}\left(\dfrac{\partial A_i}{\partial x_j}\dfrac{\partial A_k}{\partial x_l}\right) - r_{ijkl}u_{ij}A_k A_l \end{pmatrix} \quad (1b)$$

$P_i$ is the polarization vector, $u_{ij}(\mathbf{x})$ is the strain tensor ($i,j = 1 - 3$); $g_{ijkl}$ and $v_{ijkl}$ are the gradient coefficients; $f_{ijkl}$ is the forth-rank tensor of flexoelectric coupling experimentally determined for SrTiO$_3$ by Zubko et al [21], $q_{ijkl}$ is the forth-rank electrostriction tensor, $r_{ijkl}$ is the rotostriction tensor, $c_{ijkl}$ is the elastic stiffness; $\xi_{ijkl}$ is the biquadratic coupling term [22, 23]. $A_i$ are the components of the structural order parameter (OP) (see e.g. Ref. [13]). Note, that for the case of the oxygen octahedra rotation transition as an order parameter could be chosen either the rotation angle or the displacement of an appropriate oxygen atom from its cubic position [24]. The temperature dependence of the coefficients $\alpha_i$ and $\beta_i$ can be fitted with Barrett law [25], $\alpha_i(T) = \alpha_T T_q^{(E)}\left(\coth\left(T_q^{(E)}/T\right) - \coth\left(T_q^{(E)}/T_0^{(E)}\right)\right)$, $\beta_i(T) = \beta_T T_q^{(\Phi)}\left(\coth\left(T_q^{(\Phi)}/T\right) - \coth\left(T_q^{(\Phi)}/T_S\right)\right)$ for quantum paraelectrics. The higher expansion coefficients $\beta_{ij}$ and $\alpha_{ij}$ are weakly temperature dependent.

In the dielectric limit (i.e. under the absence of screening by internal carriers) the depolarization electric field components $E_i^d$ are determined self-consistently from Maxwell equation $\varepsilon_0 \varepsilon_b \dfrac{\partial E_i^d}{\partial x_i} = -\dfrac{\partial P_i}{\partial x_i}$, where $\varepsilon_0 = 8.85 \times 10^{-12}$ F/m is the universal dielectric constant, $\varepsilon_b$ is the "base" isotropic lattice permittivity, different from the ferroelectric soft mode permittivity [26, 27, 28].

It should be noted that the flexoelectric coupling could be rewritten in the form of Lifshitz invariant, $f_{ijkl}(u_{ij}\partial P_k/\partial x_l - P_k \partial u_{ij}/\partial x_l)/2$ [7, 11-12, 29], or in the short form $f_{ijkl}u_{ij}\dfrac{\partial P_k}{\partial x_l}$. Though both forms give the same equations of state, the choice between $f_{ijkl}u_{ij}\dfrac{\partial P_k}{\partial x_l}$ and Lifshitz invariant has crucial implications on the boundary conditions [12, 30].

The surface energy density has the form:
$$f_S = a_{ij}^S A_i A_j + b_{ij}^S P_i P_j \quad (1c)$$

Note, that the surface energy coefficients $a_{ij}^S$ and $b_{ij}^S$ should be termination dependent. This immediately leads to the termination dependent extrapolation lengths for the polarization and structural order parameters. It is in agreement with numerous DFT studies for termination effects in ferroelectrics [31, 32, 33, 34]. After appropriate variation, surface energy (1c) contributes to the boundary conditions $\left(2a_{ij}^S A_j + n_j v_{ijkl}\frac{\partial A_k}{\partial x_l}\right)\bigg|_S = 0$ and $\left(2b_{ij}^S P_j + n_j g_{ijkl}\frac{\partial P_k}{\partial x_l} + n_j f_{klij} u_{kl}\right)\bigg|_S = 0$, where $n_j$ is the outer normal components to the film surfaces.

The equation of state for strain and stress (generalized Hooke's law) is:

$$\sigma_{ij} = c_{ijkl} u_{kl} + f_{ijkl}\frac{\partial P_k}{\partial x_l} - q_{ijkl} P_k P_l - r_{ijkl} A_k A_l \qquad (2)$$

## 2.2. One-dimensional (1D) case

Below we consider one-dimensional (1D) case, when the structural order parameter $A_i(x_3)$ and polarization $P_i(x_3)$ depend on the distance from the interface $x_3$. Elastic fields in the epitaxial thin film clamped on a rigid substrate have the view:

$$\sigma_{33} = 0, \qquad \sigma_{13} = 0, \qquad \sigma_{23} = 0, \qquad \sigma_{12} = -q_{44} P_1 P_2 - r_{44} A_1 A_2, \qquad (3a)$$

$$\sigma_{22} = (c_{11} + c_{12})u_m + c_{12} u_{33} + f_{12}\frac{\partial P_3}{\partial x_3} - q_{12}(P_3^2 + P_1^2) - q_{11} P_2^2 - r_{12}(A_3^2 + A_1^2) - r_{11} A_2^2, \qquad (3b)$$

$$\sigma_{11} = (c_{11} + c_{12})u_m + c_{12} u_{33} + f_{12}\frac{\partial P_3}{\partial x_3} - q_{12}(P_3^2 + P_2^2) - q_{11} P_1^2 - r_{12}(A_3^2 + A_2^2) - q_{11} A_1^2 \qquad (3c)$$

$$u_{33} = -2\frac{c_{12}}{c_{11}} u_m + \frac{1}{c_{11}}\left(q_{11} P_3^2 + q_{12} P_2^2 + q_{12} P_1^2 + r_{11} A_3^2 + r_{12} A_2^2 + r_{12} A_1^2\right) - \frac{f_{11}}{c_{11}}\frac{\partial P_3}{\partial x_3}, \qquad (3d)$$

$$u_{11} = u_{22} = u_m, \qquad u_{12} = 0, \qquad (3e)$$

$$u_{13} = \frac{q_{44} P_1 P_3 + r_{44} A_1 A_3}{2c_{44}} - \frac{f_{44}}{2c_{44}}\frac{\partial P_1}{\partial x_3}, \qquad u_{23} = \frac{q_{44} P_2 P_3 + r_{44} A_2 A_3}{2c_{44}} - \frac{f_{44}}{2c_{44}}\frac{\partial P_2}{\partial x_3}. \qquad (3f)$$

We regard that the bulk material in the parent high temperature phase has m3m symmetry in order to derive Eqs.(3).

Substituting the elastic solution Eq.(3) into equations of state $\delta f_b / \delta P_i = 0$ and $\delta f_b / \delta A_i = 0$, we derived the free energy functional with renormalized coefficients:

$$\tilde{f}_b = \tilde{f}_{\mathrm{hom}} + \tilde{f}_{grad} + \tilde{f}_{electric} \qquad (4a)$$

$$\tilde{f}_{hom} \cong \beta_3^*(T,u_m)P_3^2 + \beta_{33}^* P_3^4 + \beta_{13}^*(P_1^2 + P_2^2)P_3^2 + \beta_1^*(T,u_m)(P_1^2 + P_2^2)$$
$$+ \beta_{12}^* P_1^2 P_2^2 + \beta_{11}^*(P_1^4 + P_2^4) + \xi_{11}^*(P_1^2 A_1^2 + P_2^2 A_2^2) + \xi_{33}^* P_3^2 A_3^2 + \xi_{31}^* P_3^2(A_1^2 + A_2^2)$$
$$+ \xi_{12}^*(P_1^2 A_2^2 + P_2^2 A_1^2) + \xi_{13}^*(P_1^2 + P_2^2)A_3^2 + \xi_{44}^* P_1 P_2 A_1 A_2 + \xi_{44}^*(P_1 A_1 + P_2 A_2)P_3 A_3 \quad (4b)$$
$$+ \alpha_3^*(T,u_m)A_3^2 + \alpha_{33}^* A_3^4 + \alpha_{13}^*(A_1^2 + A_2^2)A_3^2 + \alpha_1^*(T,u_m)(A_1^2 + A_2^2) + \alpha_{12}^* A_1^2 A_2^2 + \alpha_{11}^*(A_1^4 + A_2^4)$$

$$\tilde{f}_{grad} \cong \frac{g_\parallel}{2}\left[\left(\frac{\partial P_1}{\partial x_3}\right)^2 + \left(\frac{\partial P_2}{\partial x_3}\right)^2\right] + \frac{g_\perp}{2}\left(\frac{\partial P_3}{\partial x_3}\right)^2 + \frac{f_{11}q_{12}}{c_{11}}(P_1^2 + P_2^2)\frac{\partial P_3}{\partial x_3}$$
$$+ \frac{f_{44}q_{44}}{c_{44}}P_3\left(P_1\frac{\partial P_1}{\partial x_3} + P_2\frac{\partial P_2}{\partial x_3}\right) + \frac{v_{44}}{2}\left[\left(\frac{\partial A_1}{\partial x_3}\right)^2 + \left(\frac{\partial A_2}{\partial x_3}\right)^2\right] + \frac{v_{11}}{2}\left(\frac{\partial A_3}{\partial x_3}\right)^2 \quad (4b)$$
$$+ \left(\frac{f_{11}r_{12}}{c_{11}}(A_1^2 + A_2^2) + \frac{f_{11}r_{11}}{c_{11}}A_3^2\right)\frac{\partial P_3}{\partial x_3} + \frac{f_{44}r_{44}}{c_{44}}A_3\left(A_1\frac{\partial P_1}{\partial x_3} + A_2\frac{\partial P_2}{\partial x_3}\right)$$

Note, that the form of the energy (4) coincides with the one obtained by the substitution of Eq.(3) into Eq.(1b) and making a Legendre transformation $\tilde{f}_b = f_b - \sigma_{33}u_{33} - 2\sigma_{13}u_{13} - 2\sigma_{23}u_{23}$.

The expansion coefficients of the structural order parameter in Eq.(4b) are renormalized by rotostriction coupling ($r_{ij}$) and misfit strain $u_m$ that are introduced as:

$$\alpha_1^*(T,u_m) = \alpha_1(T) - \left(r_{12} + r_{11} - 2r_{12}\frac{c_{12}}{c_{11}}\right)u_m, \quad \alpha_3^*(T,u_m) = \alpha_1(T) - 2\left(r_{12} - r_{11}\frac{c_{12}}{c_{11}}\right)u_m \quad (5a)$$

$$\alpha_{11}^* = \alpha_{11} - \frac{r_{12}^2}{2c_{11}}, \quad \alpha_{12}^* = \alpha_{12} - \frac{r_{12}^2}{c_{11}}, \quad \alpha_{13}^* = \alpha_{12} - \frac{r_{11}r_{12}}{c_{11}} - \frac{r_{44}^2}{2c_{44}}, \quad \alpha_{33}^* = \alpha_{11} - \frac{r_{11}^2}{2c_{11}} \quad (5b)$$

The biquadratic coupling terms $\xi_{ij}$ are renormalized by rotostriction ($r_{ij}$) and electrostriction ($q_{ij}$) coupling:

$$\xi_{11}^* = \xi_{11} - \frac{r_{12}q_{12}}{c_{11}}, \quad \xi_{44}^* = \xi_{44} - \frac{r_{44}q_{44}}{c_{44}}, \quad \xi_{33}^* = \xi_{11} - \frac{r_{11}q_{11}}{c_{11}}, \quad (6a)$$

$$\xi_{12}^* = \xi_{21}^* = \xi_{12} - \frac{r_{12}q_{12}}{c_{11}}, \quad \xi_{13}^* = \xi_{12} - \frac{r_{11}q_{12}}{c_{11}}, \quad \xi_{31}^* = \xi_{12} - \frac{r_{12}q_{11}}{c_{11}}. \quad (6b)$$

The expansion coefficients of the polarization are renormalized by electrostriction coupling ($q_{ij}$) and misfit strain $u_m$:

$$\beta_1^*(T,u_m) = \beta_1(T) - \left(q_{12} + q_{11} - 2q_{12}\frac{c_{12}}{c_{11}}\right)u_m, \quad \beta_3^*(T,u_m) = \beta_1(T) - 2\left(q_{12} - q_{11}\frac{c_{12}}{c_{11}}\right)u_m, \quad (7a)$$

$$\beta_{11}^* = \beta_{11} - \frac{q_{12}^2}{2c_{11}}, \quad \beta_{12}^* = \beta_{12} - \frac{q_{12}^2}{c_{11}}, \quad \beta_{13}^* = \beta_{12} - \frac{q_{11}q_{12}}{c_{11}} - \frac{q_{44}^2}{2c_{44}}, \quad \beta_{33}^* = \beta_{11} - \frac{q_{11}^2}{2c_{11}}. \quad (7b)$$

The gradient coefficients ($g_{ij}$) are renormalized by the flexoelectric coupling ($f_{ij}$) as:[7]

$$g_{\parallel} = \left(g_{44} - \frac{f_{44}^2}{c_{44}}\right), \qquad g_{\perp} = \left(g_{11} - \frac{f_{11}^2}{c_{11}}\right). \qquad (7c)$$

For the short-circuited boundary conditions (i.e. when electric potential is zero at the conducting electrodes), the depolarization field is $E_3^d = -\left(P_3(x_3) - \overline{P_3}\right)/\varepsilon_0 \varepsilon_b$, where $\overline{P_3} = \frac{1}{h}\int_{-h/2}^{+h/2} dx_3 P_3(x_3)$. For the open-circuited boundary conditions (i.e. when electric displacement is zero outside the film) depolarization field is $E_3^d = -P_3(x_3)/\varepsilon_0 \varepsilon_b$. For the open-circuited boundary conditions the electric energy $\tilde{f}_{electric} = -P_i E_i - \frac{\varepsilon_0 \varepsilon_b \mathbf{E}^2}{2}$ should be modified by adding the term $D_i E_i$ [35] (see also [36]).

Variation of the bulk free energy (1b) and surface energy (1c) densities with respect to the structural order parameter $A_i(x_3)$ and polarization $P_i(x_3)$ leads to equations of state supplemented by the mixed boundary conditions

$$\left(2a_1^S A_1 + n_3 v_{44}\frac{\partial A_1}{\partial x_3}\right)\bigg|_S = 0, \quad \left(2a_1^S A_2 + n_3 v_{44}\frac{\partial A_2}{\partial x_3}\right)\bigg|_S = 0, \quad \left(2a_3^S A_3 + n_3 v_{11}\frac{\partial A_3}{\partial x_3}\right)\bigg|_S = 0 \quad (8a)$$

$$\left(2b_1^S P_1 + n_3 g_{\parallel}\frac{\partial P_1}{\partial x_3} + n_3 \frac{f_{44}}{c_{44}}\left(q_{44} P_1 P_3 + r_{44} A_1 A_3\right)\right)\bigg|_S = 0 \qquad (8b)$$

$$\left(2b_1^S P_2 + n_3 g_{\parallel}\frac{\partial P_2}{\partial x_3} + n_3 \frac{f_{44}}{c_{44}}\left(q_{44} P_2 P_3 + r_{44} A_2 A_3\right)\right)\bigg|_S = 0 \qquad (8c)$$

$$\left(\begin{array}{l} 2b_3^S P_3 + n_3 g_{\perp}\frac{\partial P_3}{\partial x_3} + 2n_3\left(f_{12} - \frac{c_{12}}{c_{11}}f_{11}\right)u_m + \\ n_3 \frac{f_{11}}{c_{11}}\left(q_{11} P_3^2 + q_{12} P_2^2 + q_{12} P_1^2 + r_{11} A_3^2 + r_{12} A_2^2 + r_{12} A_1^2\right) \end{array}\right)\bigg|_S = 0 \qquad (8d)$$

The components of the outer normal to the film surfaces $x_3 = \mp h/2$ are $n_3 = \pm 1$ correspondingly. Hereinafter we assumed that the tensors $a_{ij}^S \equiv \delta_{ij} a_i^S$ and $b_{ij}^S \equiv \delta_{ij} b_i^S$ are diagonal and, more important, chosen the same for both film surfaces for simplicity only. The assumption leads to the symmetric (with respect to the interfaces $x_3 = \mp h/2$) boundary conditions (8) and thus to the symmetric solutions. In general case different interface energy should lead to the asymmetric boundary conditions.

It may seem that one can neglect all nonlinear terms in the boundary conditions (8b)-(8d) for estimations, but exactly some of the terms couple polarization with structural order parameter and thus can strongly affect phase diagrams and finite size effects in thin multiferroic films.

## 3. Approximate analytical solution in decoupling approximation

### 3.1. Structural order parameter components

Numerical analyses showed that in the structural phase (e.g. octahedrally tilted phase $\alpha_1^* < 0$), polarization value, orientation and corresponding derivatives weakly influence on the structural order parameters $A_i$ in such materials as perovskites $SrTiO_3$, $CaTiO_3$, $EuTiO_3$. For the case all polarization-dependent terms can be omitted in equations for the structural order parameters, and, in this *decoupling approximation*, approximate analytical solution acquires the form:[37]

$$A_1(x_3) = A_b \sqrt{\frac{2m}{1+m}} \, \text{sn}\left(\frac{x_3}{L_1^A(T, u_m)\sqrt{1+m}} + c_0 \bigg| m\right), \quad A_2(x_3) = 0. \qquad (9)$$

Here sn($u|m$) is the elliptical sine function. For the case of identically zero or negligibly small $A_3$, the amplitude $A_b = \sqrt{-\alpha_1^*/2\alpha_{11}^*}$ and characteristic length $L_1^A(T, u_m) = \sqrt{\dfrac{-v_{44}}{2\alpha_1^*(T, u_m)}}$, which should be positive in the structural phase. For the case $A_3 \neq 0$, one can use the approximations

$$A_b \approx \sqrt{-\frac{2\alpha_1^*(T, u_m) - (\alpha_3^* \alpha_{13}^*/\alpha_{33}^*)}{4\alpha_{11}^* - (\alpha_{13}^{*2}/\alpha_{33}^*)}} \quad \text{and} \quad L_1^A(T, u_m) \approx \sqrt{\frac{-v_{44}}{2\alpha_1^*(T, u_m) - (\alpha_3^* \alpha_{13}^*/\alpha_{33}^*)}},$$

which are self-consistent with Eq.(12) under the condition $\alpha_{13}^* A_1^2 + \alpha_3^* < 0$.

The constants $m$ and $c_0$ should be determined from the symmetric boundary condition (8) at the film surfaces $x_3 = \pm h/2$. Symmetry of the boundary conditions gives $c_0 = K(m)/2$, where K($m$) is the complete elliptic integrals of the first kind [38]. Similar solution is obtained with the substitution of subscripts: 1↔2.

At $m \to 0$ the term $A_1(x_3)$ vanishes, corresponding to the onset of the structural phase transition into either cubic phase or another structural phase. The condition of $A_1(x_3)$ vanishing is given by equation

$$\cos\left(\frac{h}{2L_1^A(T, u_m)}\right) = \frac{\lambda_{S1}}{L_1^A(T, u_m)} \sin\left(\frac{h}{2L_1^A(T, u_m)}\right), \qquad (10)$$

where the structural order parameter extrapolation length $\lambda_{S1} = \dfrac{v_{44}}{2a_1^S}$ is introduced. The way of derivation of expression (10) is explained in Ref.[ 39 ]. The *critical film thickness* $h_{cr}^A = 2L_1^A(T, u_m) \arctan\left(\dfrac{L_1^A(T, u_m)}{\lambda_{S1}}\right)$ below which the structural order parameters $A_{1,2}$ disappear can be determined from Eq.(10) at given temperature $T$, misfit strain $u_m$ and extrapolation length $\lambda_{S1}$

The solution different from (9) is

$$A_1(x_3) = A_2(x_3) = A(x_3) \approx A_c \sqrt{\frac{2m}{1+m}} \, \text{sn}\left( \frac{x_3}{L_1^A(T,u_m)\sqrt{1+m}} + \frac{K(m)}{2} \bigg| m \right), \quad (11)$$

where the amplitude $A_c = \sqrt{-\alpha_1^*/(2\alpha_{11}^* + \alpha_{12}^* + \alpha_{13}^*)}$ for identically zero or negligibly small $A_3$. For $A_3 \neq 0$ the amplitude $A_b \approx \sqrt{-\frac{2\alpha_1^*(T,u_m) - (\alpha_3^* \alpha_{13}^*/\alpha_{33}^*)}{(4\alpha_{11}^* + 2\alpha_{12}^* + 2\alpha_{13}^*) - 2(\alpha_{13}^{*2}/\alpha_{33}^*)}}$, which is self-consistent with Eq.(12) under the condition $2\alpha_{13}^* A_1^2 + \alpha_3^* < 0$. Length $L_1^A(T,u_m)$ is the same as in Eq.(9). At $m \to 0$ $A_1(x_3)$ vanishes, which means the structural phase transition. The boundary of the structural phase $A_1(x_3) = A_2(x_3) = A(x_3)$ is also given by Eq.(10).

Typically the gradient coefficient $v_{44} \gg v_{11}$ [40]. As a result, the characteristic lengths $L_1^A(T,u_m) = L_2^A(T,u_m)$, which determine the gradient scale of the structural order parameter components $A_1$ and $A_2$, are typically higher than lattice constant(s), but the length $L_3^A(T,u_m) = \left| \frac{v_{11}}{2\alpha_3^*(T,u_m)} \right|^{1/2}$ that determines the gradient scale of the structural order parameter component $A_3$, is (much) smaller than the lattice constant, thus approximate analytical solution for $A_3$ is:

$$A_3(x_3) \approx \begin{cases} \sqrt{-\dfrac{\alpha_{13}^*(A_1^2(x_3) + A_2^2(x_3)) + \alpha_3^*}{2\alpha_{33}^*}}, & \alpha_{13}^*(A_1^2 + A_2^2) + \alpha_3^* < 0 \\ 0, & \alpha_{13}^*(A_1^2 + A_2^2) + \alpha_3^* > 0 \end{cases} \quad (12)$$

Note, that the accuracy of approximate solution (12) is the best for the case $a_3^S = 0$, when $\frac{\partial A_3}{\partial x_3} = 0$ at the film surfaces $x_3 = \mp h/2$. To be self-consistent hereinafter we regard $a_3^S = 0$. The boundary of the structural phase with $A_3(x_3) \neq 0$ is given by equation $\alpha_{13}^*(A_1^2(x_3) + A_2^2(x_3)) + \alpha_3^* = 0$.

### 3.2. Polarization components

Relatively simple results for polarization can be derived for the **open-circuited electrical boundary conditions**, since $P_3(x_3)$ appeared negligibly small due to the strong depolarization field. Moreover, corresponding spatial scale, $L_3^P(T,u_m) = \left| \frac{g_\perp}{2\beta_3^*(T,u_m) + 1/\varepsilon_0 \varepsilon_b} \right|^{1/2}$, appeared ~0.1 nm, i.e. significantly smaller than the lattice constant (~0.4 nm) of considered perovskites. Thus putting $P_3(x_3) \approx 0$, we should solve two coupled equations for polarization components $P_1$ and $P_2$ only.

In the ***tetragonal-like*** structural, $A_1(x_3) = A_2(x_3) \equiv A(x_3)$, $P_1(x_3) = P_2(x_3) \equiv P(x_3)$, where $P(x_3)$ obeys equation:

$$\gamma(x_3, A, A_3)P + (4\beta_{11}^* + 2\beta_{12}^*)P^3 - g_\| \frac{\partial^2 P}{\partial x_3^2} = E_\|^{FR}(A, A_3) \tag{13}$$

The function $\gamma(x_3, A, A_3) = 2\beta_1^* + \xi_{44}A^2 + 2((\xi_{11}^* + \xi_{12}^*)A^2(x_3) + \xi_{13}^* A_3^2(x_3))$ and flexo-roto field $E_\|^{FR}(A, A_3) = \frac{f_{44} r_{44}}{c_{44}} \frac{\partial(A_3 A)}{\partial x_3}$ depends on the structural parameter components. The structural parameter components $A_1 = A_2 \equiv A(x_3)$ and $A_3(x_3)$ are given by Eq.(9) and (12) correspondingly. To be consistent with approximation (12) we should put $\frac{\partial(A_3 A)}{\partial x_3} \approx A_3 \frac{\partial A}{\partial x_3}$. The boundary conditions to Eq.(13) is $\left( P \pm \lambda_{P1} \frac{\partial P}{\partial x_3} \pm \frac{f_{44} r_{44}}{2 b_1^S c_{44}} A A_3 \right)\bigg|_{x_3 = \pm h/2} = 0$, where $\lambda_{P1} = \frac{g_\|}{2 b_1^S}$ is extrapolation length of the polarization.

In the ***monoclinic-like*** structural phase, $A_2(x_3) = 0$ and $A_1(x_3) \neq 0$ (or $A_1(x_3) = 0$ and $A_2(x_3) \neq 0$), $P_1(x_3) \neq 0$ and $P_2(x_3) = 0$ (or $P_1(x_3) = 0$ and $P_2(x_3) \neq 0$), where $P_1(x_3)$ obeys equation:

$$\gamma(x_3, A_1, A_3)P_1 + 4\beta_{11}^* P_1^3 - g_\| \frac{\partial^2 P_1}{\partial x_3^2} = E_1^{FR}(A_1, A_3) \tag{14}$$

The function $\gamma(x_3, A_1, A_3) = 2(\beta_1^* + \xi_{11}^* A_1^2(x_3) + \xi_{13}^* A_3^2(x_3))$ and the flexo-roto field $E_1^{FR}(A_1, A_3) = \frac{f_{44} r_{44}}{c_{44}} \frac{\partial(A_3 A_1)}{\partial x_3}$ are introduced. The structural parameter components $A_1$ and $A_3$ are given by Eq.(10) and (12) correspondingly. To be consistent with approximation (12) we should put $\frac{\partial(A_3 A_1)}{\partial x_3} \approx A_3 \frac{\partial A_1}{\partial x_3}$. The boundary conditions to Eq.(14) are $\left( P_1 \pm \lambda_{P1} \frac{\partial P}{\partial x_3} \pm \frac{f_{44} r_{44}}{2 b_1^S c_{44}} A_1 A_3 \right)\bigg|_{x_3 = \pm h/2} = 0$.

The characteristic lengths $L_1^P(T, u_m) = L_2^P(T, u_m) = \left| \frac{g_\|}{2\beta_1^*(T, u_m)} \right|^{1/2}$, which determine the gradient scale of the polarization components $P_1$ and $P_2$, are typically higher than lattice constant(s). The fact allows us to solve Eqs.(13), (14) along with extrapolation length-dependent boundary conditions. Solution of Eqs.(14) can be found numerically, but approximate analytics is possible for determination of the ferroelectric and polar phase boundaries [as argued in the **Supplementary Materials**]. The following transcendental equation determines the in-plane ferroelectric phase boundary in dependence on film thickness $h$, misfit strain $u_m$ and temperature $T$:

$$\cos\left(\int_0^{h/2}\sqrt{\frac{\gamma(x,A_i(x))}{-g_\parallel}}dx\right)=\lambda_{P1}\sqrt{\frac{\gamma(h/2,A_i(h/2))}{-g_\parallel}}\sin\left(\int_0^{h/2}\sqrt{\frac{\gamma(x,A_i(x))}{-g_\parallel}}dx\right) \quad (15)$$

The boundary (15) depends on the extrapolation length $\lambda_P$. The function $\gamma(x, A_i(x))$ is different for the structural phases $A_1(x_3) = A_2(x_3) \equiv A(x_3)$ and $A_2(x_3) = 0$, $A_1(x_3) \neq 0$ (compare Eq.(13) and (14) correspondingly). The critical thickness $h_{cr}^P$, below which the *switchable* ferroelectric polarization components disappear can be determined from Eq.(15) at given temperature $T$, misfit strain $u_m$ and extrapolation length $\lambda_P$.

The boundary of the local polar phase induced by the flexo-roto field is diffuse and can be estimated from the condition $E_{FR}(A_1, A_2, A_3) \approx 0$ indicating the flexo-roto field vanishing far from the film surfaces. For the **short-circuited electric boundary conditions** out-of-plane polarization component appears and it is coupled with corresponding flexo-roto field component $E_3^{FR} = \frac{\partial}{\partial x_3}\left(\frac{f_{11}r_{12}}{c_{11}}(A_1^2 + A_2^2) + \frac{f_{11}r_{11}}{c_{11}}A_3^2\right)$ (see Eq.(A.2g) in **Suppl. Mat.**). For the case of short-circuited conditions the critical thickness of the out-of-plane switchable ferroelectric polarization appearance can be determined from equation:

$$\cosh\left(\int_0^{h/2}\sqrt{\frac{\gamma(x)}{g_\perp}}dx\right) + \lambda_{P3}\sqrt{\frac{\gamma(h/2)}{g_\perp}}\sinh\left(\int_0^{h/2}\sqrt{\frac{\gamma(x)}{g_\perp}}dx\right) = 0, \quad (16)$$

where $\gamma(x) = 2\beta_3^* + 2(\xi_{33}^* A_3^2 + \xi_{31}^*(A_1^2 + A_2^2)) + \frac{1}{\varepsilon_0\varepsilon_b}$ and typically $\gamma(x) \approx \frac{1}{\varepsilon_0\varepsilon_b}$.

### 4. Phase diagrams

*4.1. Approximate expressions for the phase boundaries*

In order to calculate phase diagrams of thin films, of interest are the surface energy coefficients $a_i^S \geq 0$ and $b_i^S \geq 0$. These parameters uniquely determine corresponding extrapolation lengths

$$\lambda_{P1} = \frac{g_\parallel}{2b_1^S}, \quad \lambda_{P3} = \frac{g_\perp}{2b_3^S}, \quad \lambda_{S1} = \frac{v_{44}}{2a_1^S}, \quad \lambda_{S3} = \frac{v_{11}}{2a_3^S}, \quad (17)$$

The physical meaning of these parameters is discussed in detail in Refs.[17, 18]. Since the value and temperature dependence of the surface energy coefficients (and thus extrapolation lengths) are priory unknown for a given ferroic material, we can vary it in the actual range of 1 – 100 lattice constants.

Analytical calculations, based on the free energy (1) minimization by direct variational method with trial functions taken in the form of elliptic (when depolarization field is absent) or hyperbolic (when depolarization field is present) functions [39], after the integration on $x_3$ over the film thickness $h$, leads to the renormalization of the expansion coefficients for the structural order parameter components:

$$\alpha_3^R(T,u_m,h) \approx \alpha_3^*(T,u_m) + \frac{\pi^2 v_{11}}{\pi^2 h \lambda_{S3} + 2h^2} \equiv \alpha(T) + \frac{\pi^2 v_{11}}{\pi^2 h \lambda_{S3} + 2h^2} - 2\left(r_{12} - r_{11}\frac{c_{12}}{c_{11}}\right)u_m, \quad (18a)$$

$$\alpha_1^R(T,u_m,h) \approx \alpha_1^*(T,u_m) + \frac{\pi^2 v_{44}}{\pi^2 h \lambda_{S1} + 2h^2} \equiv \alpha(T) + \frac{\pi^2 v_{44}}{\pi^2 h \lambda_{S1} + 2h^2} - \left(r_{12} + r_{11} - 2r_{12}\frac{c_{12}}{c_{11}}\right)u_m, \quad (18b)$$

The expansion coefficients for the in-plane and out-of-plane polarization components are:

$$\beta_1^R(T,u_m,h) \approx \beta_1^*(T,u_m) + \frac{\pi^2 g_\parallel}{\pi^2 h \lambda_{P1} + 2h^2} \equiv \beta(T) + \frac{\pi^2 g_\parallel}{\pi^2 h \lambda_{P1} + 2h^2} - \left(q_{12} + q_{11} - 2q_{12}\frac{c_{12}}{c_{11}}\right)u_m, \quad (18c)$$

$$\beta_3^R(T,u_m,h) \approx \beta_3^*(T,u_m) + \frac{g_\perp}{(\lambda_{P3} + L_\perp)h} \equiv \beta(T) + \frac{g_\perp}{(\lambda_{P3} + L_\perp)h} - 2\left(q_{12} - q_{11}\frac{c_{12}}{c_{11}}\right)u_m. \quad (18d)$$

Note that Eq.(18b) is valid for the *short-circuited boundary conditions* and $L_\perp = \sqrt{\varepsilon_0 \varepsilon_b g_\perp} \sim 0.2$ nm. For the *open-circuited boundary conditions* the renormalized coefficient

$$\beta_3^R(T,u_m,h) \approx \beta_3^*(T,u_m) + \frac{1}{2\varepsilon_0 \varepsilon_b} + \frac{\pi^2 g_\perp}{\pi^2 h \lambda_{P3} + 2h^2} \quad (18e)$$

is typically positive due to the big positive additive $(2\varepsilon_0 \varepsilon_b)^{-1}$ originated from the depolarization field.

Using Eqs.(18) allows the boundary between the structural or polar phases and "high temperature" phase with zero $A_{1,2}$, $A_3$, $P_{1,2}$ and $P_3$ to be estimated from conditions $\alpha_1^R(T,u_m,h)=0$, $\alpha_3^R(T,u_m,h)=0$, $\beta_1^R(T,u_m,h)=0$ and $\beta_3^R(T,u_m,h)=0$ correspondingly. The equilibrium boundaries between different structural and/or polar phases should be determined numerically from the condition of the corresponding energies equality for a given misfit strain $u_m$.

### 4.2. Simulation results
#### 4.2.1. Phase diagrams in coordinates "temperature –SrTiO$_3$ film thickness"

The phase diagrams in coordinates "temperature –SrTiO$_3$ film thickness" are shown in **Figs 2-5.** All diagrams are calculated for the *short-circuited electric boundary conditions*, which are the most favorable for $P_3$ appearance and most typical for thin film applications as e.g. corresponding to the thin film placed between conducting electrodes (planar capacitor). Different plots (a)-(d) on each diagram are calculated for different misfit strains $u_m = 0$, -0.5, -2, 1 %. The phase labels indicate nonzero components of the structural order parameter and ferroelectric polarization components. All polar phases correspond to switchable ferroelectric polarization, while the flexo-roto fields (proportional to the gradients of the structural parameter components) can induce improper polarization components, which will be demonstrated in the next section. For instance, designation $P_1P_2A_3$-phase means the phase with nonzero ferroelectric polarization components $P_1$, $P_2$ and the structural order parameter $A_3$; $A_3$-phase means the phase with nonzero structural order parameter $A_3$; HT-phase means the parent phase with identically zero $A_i$ and $P_i$ entire the film. Solid lines denote the phase transitions between the ferroelectric, structural and HT phases. Phases with improper (and thus non-switchable) polarization components induced by the flexo-roto fields are designed as $IP_i$. For instance, an improper $IP_3$-phase can be induced by the flexo-roto field component $E_3^{FR} = \frac{\partial}{\partial x_3}\left(\frac{f_{11}r_{12}}{c_{11}}\left(A_1^2 + A_2^2\right) + \frac{f_{11}r_{11}}{c_{11}}A_3^2\right)$ (see Eq.(A.2g) in **Suppl. Mat.**) as well as by inhomogeneity in the boundary conditions $2\left(f_{12} - \frac{c_{12}}{c_{11}}f_{11}\right)u_m$ (see Eq.(A.3c) in **Suppl. Mat.**) and likely to be observed experimentally when the flexoelectric coupling is strong. The term $2\left(f_{12} - \frac{c_{12}}{c_{11}}f_{11}\right)u_m$ leads to the transformation of the HT-phase into the $IP_3$-phase for nonzero $u_m$. Similarly, phases with surface-induced structural order parameter components are designed as $IA_i$. For instance, designation $A_3P_3$-$IA_1P_2$ phase means the phase with nonzero ferroelectric polarization $P_3$ and spontaneous structural parameter $A_3$, and with components $A_1$ and improper polarization $P_2$ induced by the surface influence. Let us underline that improper (surface) *I-phases* primary appear due to the flexo-roto coupling.

For the case of zero misfit strain ($u_m = 0$), which corresponds to the film on a matched substrate, phase diagram contains the maximal amount of phases including the ferroelectric-structural $P_1P_2A_3$-phase (see plots (a) in **Figs 2-5**). Compressive (i.e. negative) misfit strain ($u_m < -0.2\%$) leads to the appearance of out-of-plane $A_3$-phases and $P_3$-phases as well as to the simultaneous disappearance of in-plane $A_{1,2}$-phases and $P_{1,2}$-phases (see plots (b,c) in **Figs 2-5**). Tensile (i.e. positive) misfit strain ($u_m > 0.05\%$) leads to the emergence of in-plane $A_{1,2}$-phases and $P_{1,2}$-phases as well as to the simultaneous disappearance of out-of-plane $A_3$-phases and $P_3$-phases (see plots (d) in **Figs 2-5**). The conclusion regarding the possibility to trigger between in-plane and out-of-plane components of spontaneous polarization and structural order parameter by the sign of misfit strain is in agreement with the phase diagram calculated by Pertsev et al. [13]. Note, that the phase diagram [13] is independent on the film thickness, since it was calculated without taking into account the gradient effects (and so without *flexoelectric coupling*). The gradient effects can be minimized in the limiting case $\lambda_{P,S} \to \infty$ (i.e. $a_i^S = b_i^S = 0$), that corresponds to the so-called "natural" boundary conditions, when the surface energy is independent on the structural order parameter and polarization and all gradients are zero inside a homogeneous film. The natural boundary conditions correspond to the constant solution, $P_i(x_3) = P_{i0}$ and $A_i(x_3) = A_{i0}$, independent of the film thickness. For the case of six constants, $P_{i0}$ and $A_{i0}$ are determined from the minimum of the bulk density (4b), and we checked that our results reproduce those in [13].

The gradient and flexoelectric effects contribution to the phase diagrams inevitably appears for finite extrapolation lengths, at that the values of critical thicknesses of the structural order parameter ($h_{cr}^A$) and polarization ($h_{cr}^P$), phase boundaries position and amount of phases strongly depend on the value of polar ($\lambda_P$) and structural ($\lambda_S$) extrapolation lengths and relations between them. In particular, the critical thicknesses $h_{cr}^A$ and $h_{cr}^P$ decrease with extrapolation length increase. To demonstrate the fact, **Figs 2-5** are calculated for different extrapolation lengths: small $\lambda_{P,S}$ ~lattice constant, high $\lambda_{P,S}$ >>lattice constant, equal $\lambda_S = \lambda_P$ and different $\lambda_S \neq \lambda_P$).

The phase diagrams in **Figure 2** are calculated for equal small extrapolation lengths $\lambda_P = \lambda_S = 0.4$ nm. Phase boundaries are strongly dependent on the film thickness and temperature for the case. The regions of the structural and polar phases increase with the film thickness increase. The boundary of parent HT-phase is almost independent on the film thickness. The critical thicknesses $h_{cr}^A$ and $h_{cr}^P$ are maximal for the case of small extrapolation lengths, but strongly dependent on misfit strain. For most cases $h_{cr}^A \leq h_{cr}^P$.

The phase diagrams in **Figure 3** are calculated for high enough equal extrapolation lengths $\lambda_P = \lambda_S = 40$ nm. Some phase boundaries are close to the horizontal lines for the case. In the sense the results tends to the limiting case $\lambda_{P,S} \to \infty$. However size effect is still pronounced at least for small misfits (see plot (a)).

The phase diagrams in **Figure 4** and **5** are calculated for different extrapolation lengths $\lambda_P \neq \lambda_S$. The cases $\lambda_P >> \lambda_S$ and $\lambda_P << \lambda_S$ look quite different (compare **Figs.4** and **5**). In the case $\lambda_P >> \lambda_S$ ferroelectric phases exist in the essentially wider temperature range than in the case $\lambda_P << \lambda_S$. As anticipated corresponding phase boundaries are close to the horizontal lines if corresponding extrapolation length is much higher than the lattice constant.

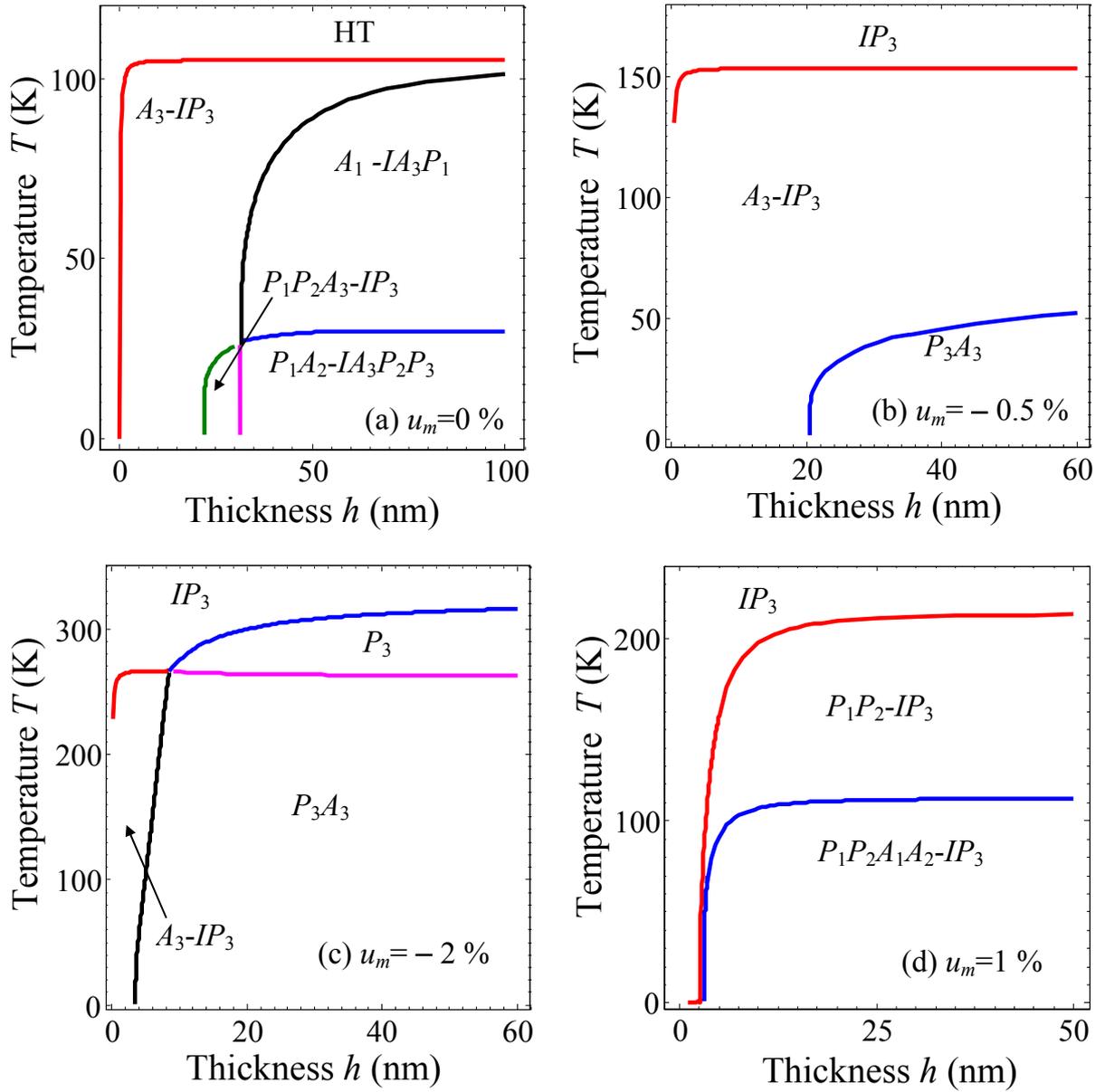

**Figure 2.** Phase diagrams in coordinates "temperature –SrTiO$_3$ film thickness". Extrapolation lengths are equal and small: $\lambda_P$=0.4 nm, $\lambda_S$=0.4 nm. Misfit strain $u_m$ = 0, -0.5, -2, 1 % (plots (a)-(d)).

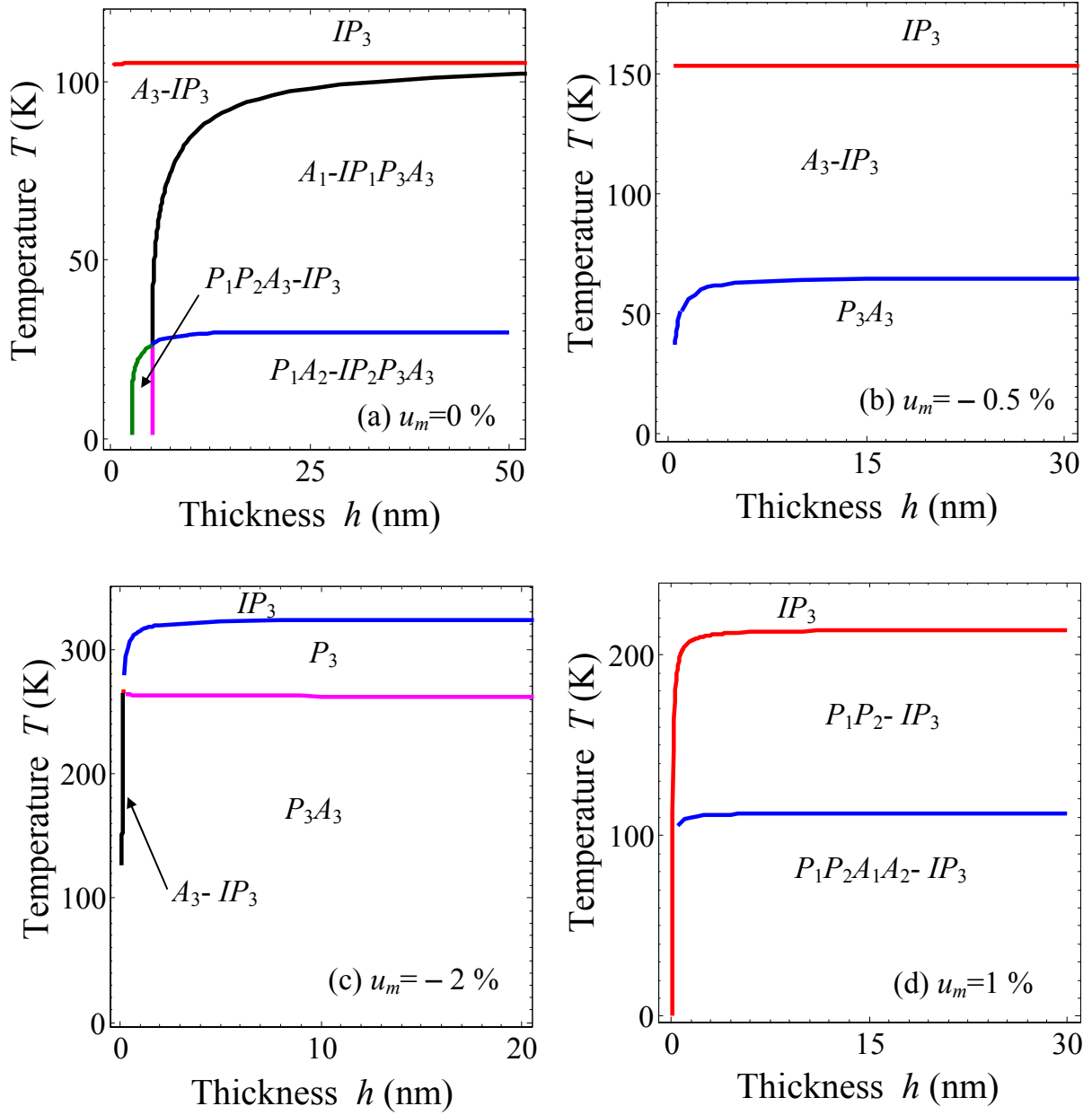

**Figure 3.** Phase diagrams in coordinates "temperature – SrTiO$_3$ film thickness". Extrapolation lengths are equal and high enough: $\lambda_P$ =40 nm, $\lambda_S$ =40 nm. Misfit strain $u_m$ = 0, -0.5, -2, 1 % (plots (a)-(d)).

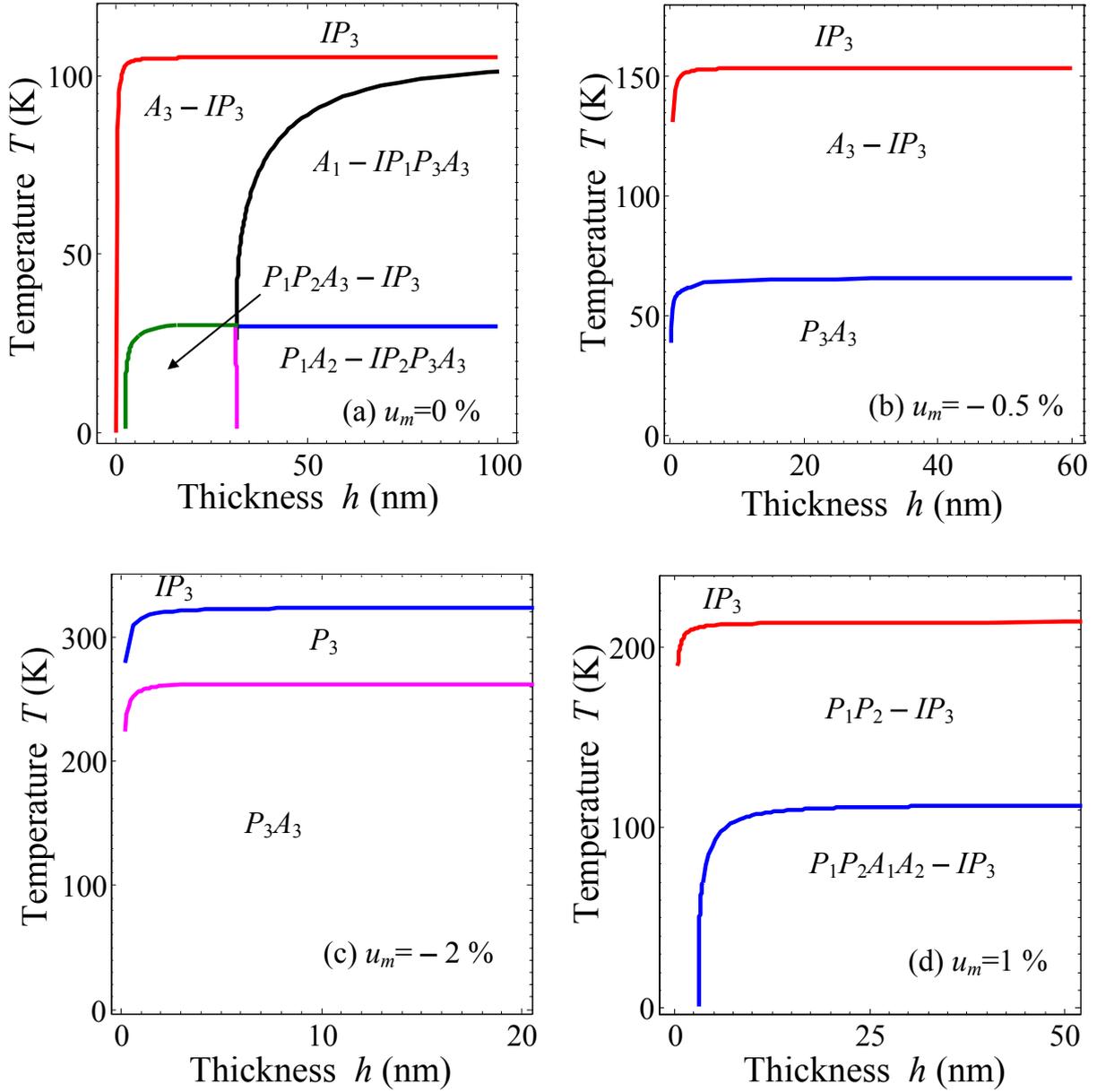

**Figure 4.** Phase diagrams in coordinates "temperature – SrTiO$_3$ film thickness". Extrapolation lengths are different: $\lambda_P$ =40 nm, $\lambda_S$ =0.4 nm. Misfit strain $u_m$ = 0, -0.5, -2, 1 % (plots (a)-(d)).

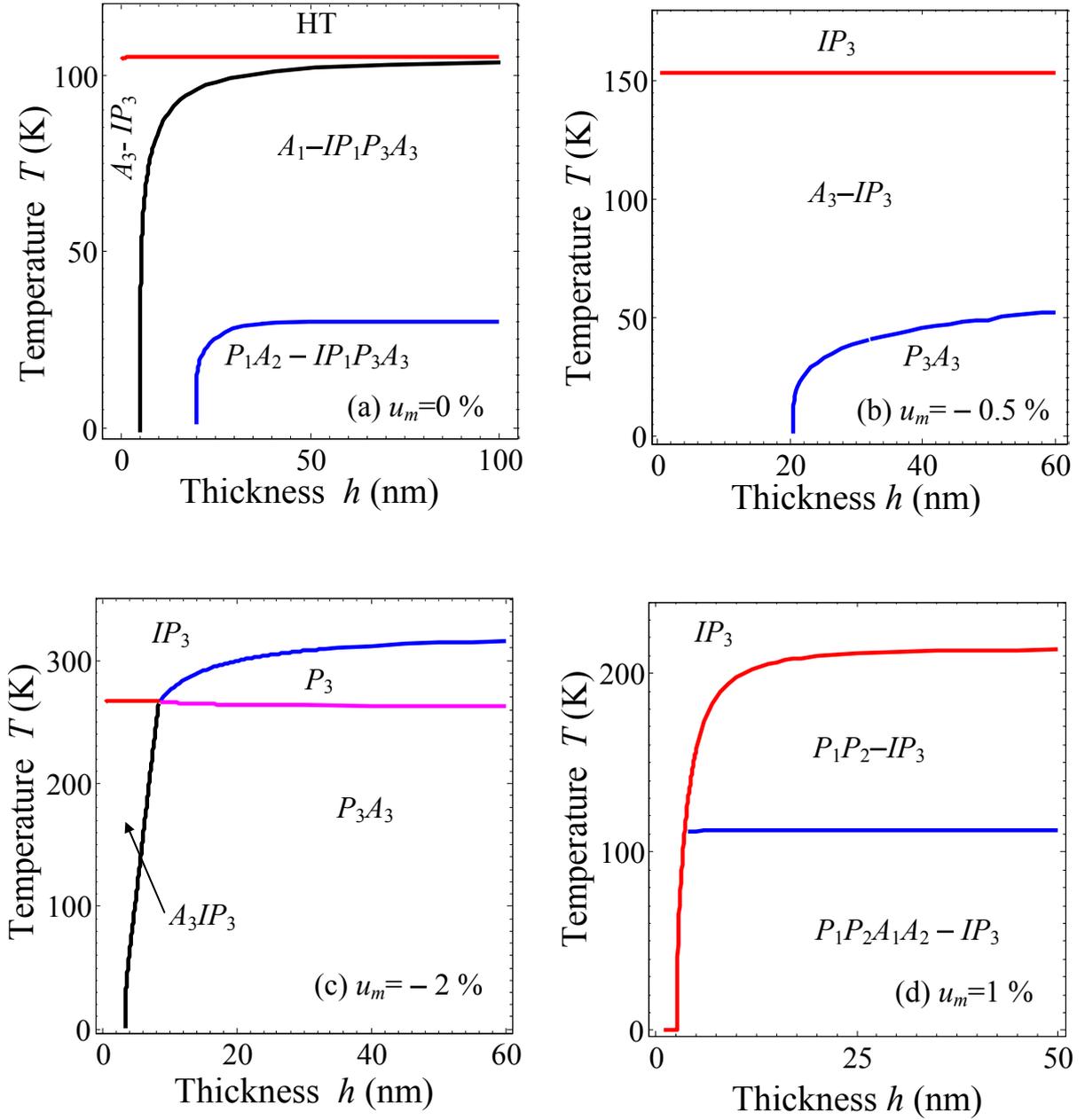

**Figure 5.** Phase diagrams in coordinates "temperature – SrTiO$_3$ film thickness". Extrapolation lengths are different: $\lambda_P$ =0.4 nm, $\lambda_S$ =40 nm. Misfit strain $u_m$ = 0, -0.5, -2, 1 % (plots (a)-(d)).

*4.2.2. Polarization and structural parameter profiles in different phases*

In order to demonstrate the difference between ferroelectric and improper polarization, spontaneous and induced structural parameter components we presented polarization and structural parameter profiles in different phases (see **Figs 6-8**).

Temperature evolution of the structural parameter components and polarization profiles in short-circuited thin (*h*=10 nm, **Figs 6**) and more thick (*h*=20 nm, **Figs 7**) SrTiO$_3$ films was calculated for zero, negative and positive misfit strains ($u_m$ =+1, +0.1, 0, –0.1, –0.5,–2 %). It is seen from the figures that the profile of the structural parameter across the film is almost symmetric with respect to the film center (small asymmetry appears only for high negative misfit strain as shown in **Fig.7e**). A spontaneous structural parameter is nonzero entire the film, maximal in the center $z = 0$ and minimal near the surfaces $z = \pm h/2$ (see e.g. $A_3$ distributions in **Figs.6a** and **6c**, $A_{1,2}$ distributions in **Figs.6e**). Improper structural parameter is nonzero only in the vicinity of film surfaces $z = \pm h/2$ as induced by the spatial confinement (see e.g. $A_3$ distributions in **Figs.7a** and **7c**). Spontaneous ferroelectric polarization distribution can be symmetric if they are not affected by the coupling with the structural parameter (see e.g. $P_{1,2}$ distributions in **Fig.6b** and **6f**, $P_1$ distributions in **Fig.7b**) or asymmetric if they are affected by the coupling with corresponding $A_i$ (see $P_3$ distributions in **Fig.6d** and **7e**). For the symmetric boundary conditions (i.e. $\lambda_P(h/2) = \lambda_P(-h/2)$, $\lambda_S(h/2) = \lambda_S(-h/2)$), improper polarization distribution is always antisymmetric inside the film (odd with respect to $z \to -z$), zero in the central part of the film and maximal near the surfaces, since it is induced either by the flexo-roto fields $E_1^{FR} = \dfrac{f_{44} r_{44}}{c_{44}} \dfrac{\partial (A_3 A_1)}{\partial x_3}$,

$E_2^{FR} = \dfrac{f_{44} r_{44}}{c_{44}} \dfrac{\partial (A_3 A_2)}{\partial x_3}$ , $E_3^{FR} = \dfrac{\partial}{\partial x_3} \left( \dfrac{f_{11} r_{12}}{c_{11}} (A_1^2 + A_2^2) + \dfrac{f_{11} r_{11}}{c_{11}} A_3^2 \right)$ (see Eq.(A.2d-g)) or by the

inhomogeneities $\pm \dfrac{f_{44}}{c_{44}} r_{44} A_1 A_3$, $\pm \dfrac{f_{44}}{c_{44}} r_{44} A_2 A_3$ and $\pm 2 \left( f_{12} - \dfrac{c_{12}}{c_{11}} f_{11} \right) u_m \pm \dfrac{f_{11}}{c_{11}} \left( r_{11} A_3^2 + r_{12} A_2^2 + r_{12} A_1^2 \right)$ in

the boundary conditions (A.3) for polarization. Signs "+" and "-" in the boundary conditions correspond to different film surfaces, so the inhomogeneous terms in the boundary conditions can induce only antisymmetric improper polarization. The flexo-roto field is antisymmetric with respect to $z \to -z$, small in the central part of the film, where the gradients of the structural parameter are small, and maximal near the surfaces, where the gradients of the structural parameter are the highest. Penetration depth of the improper phases (both polar and structural) can reach several nm from the film surfaces.

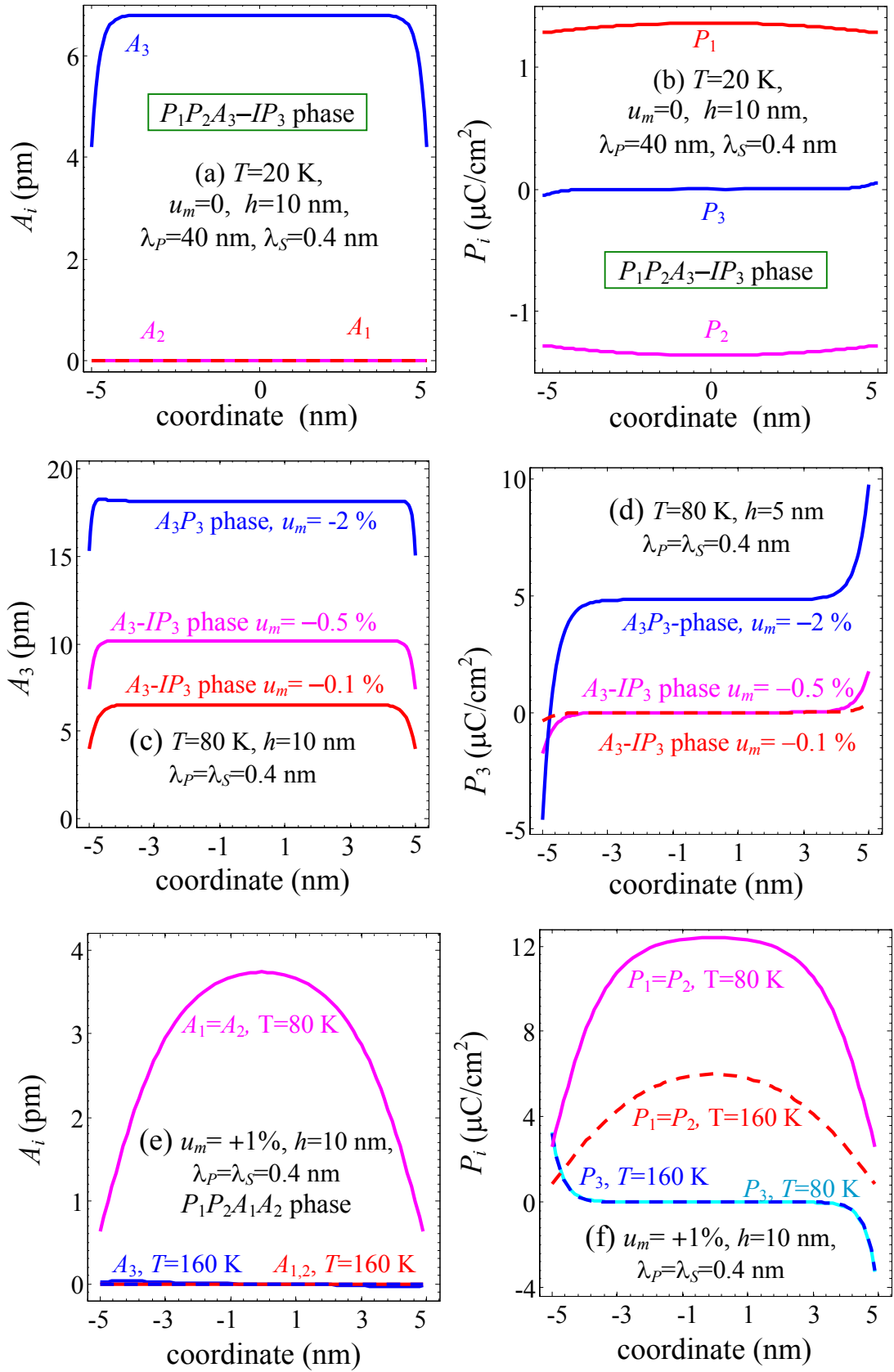

**Figure 6.** Temperature evolution of the structural parameter components and polarization profiles in short-circuited thin SrTiO$_3$ films ($h$=10 nm) calculated for different misfit strains (see numbers near the curves) and extrapolation lengths (listed at the legends inside the plots). Temperature $T$=20K (a, b); 80 K (c,d); 80 and 160 K (e,f).

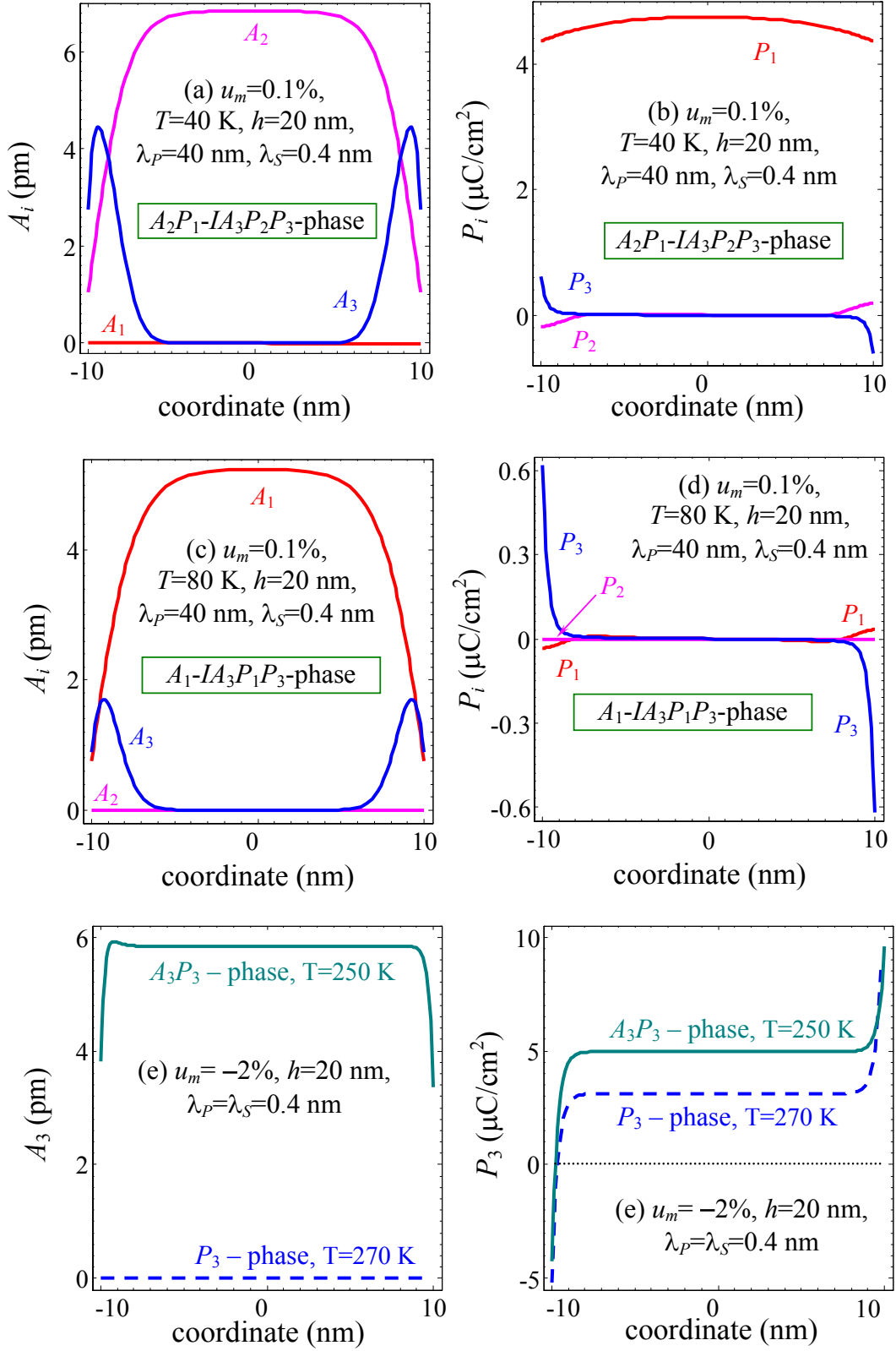

**Figure 7.** Evolution of the structural order parameter components and polarization profiles in short-circuited thicker SrTiO3 films ($h$=20 nm) with temperature increase calculated for different misfit strains (see numbers near the curves) and extrapolation lengths (listed at the legends inside the plots). Temperature $T$=40K (a, b); 80 K (c,d); 250 and 270 K (e,f).

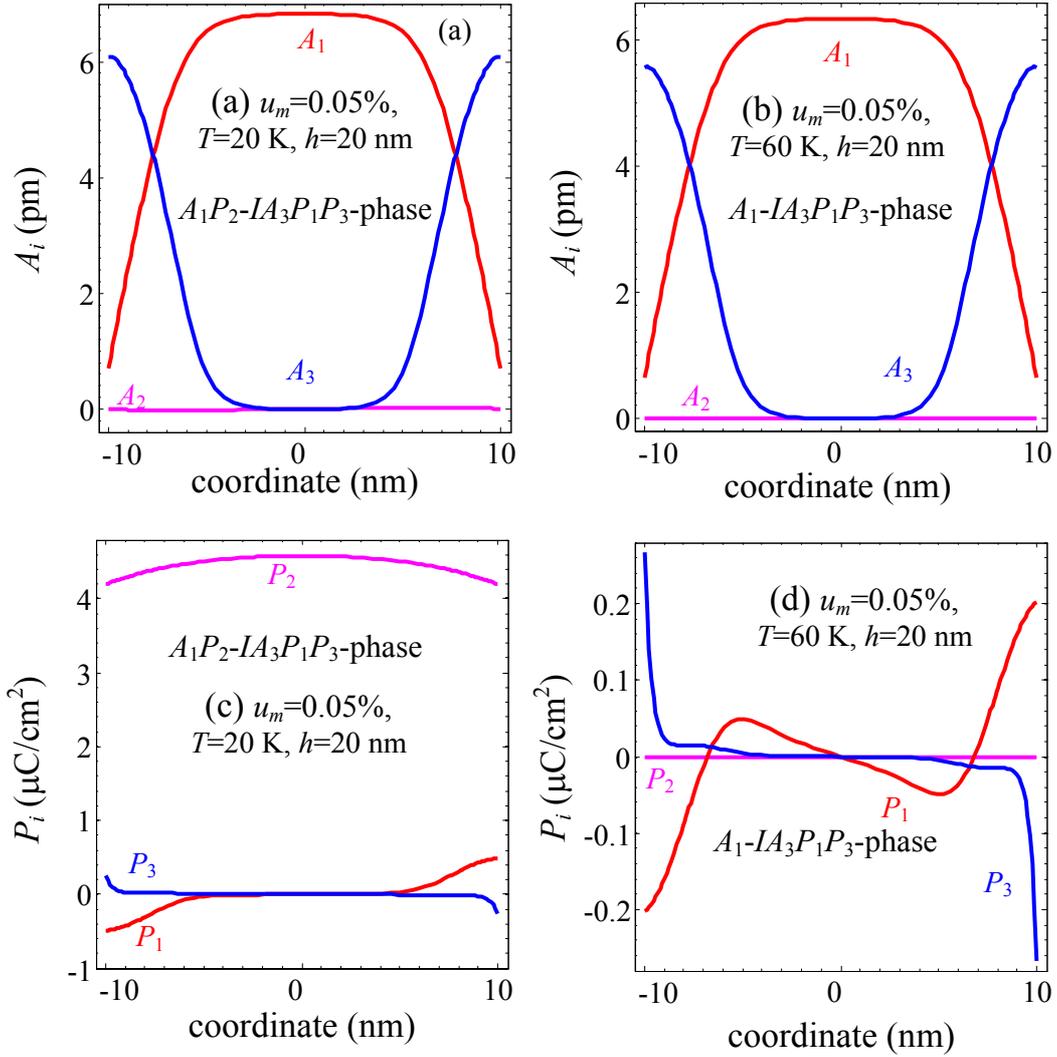

**Figure 8.** Distributions of structural order parameters (a, b), polarization (c, d) components in a short-circuited SrTiO$_3$ film of thickness $h$ = 20 nm. Misfit strain $u_m$= 0.05 % and temperature $T$ = 20 K (a, c, e) and 60 K (b, d, f). Extrapolation lengths are $\lambda_{P1}(h/2) = \lambda_{P1}(-h/2)$ = 40 nm, $\lambda_{P3}(h/2) = \lambda_{P3}(-h/2)$ = 40 nm, $\lambda_{S1}(h/2) = \lambda_{S1}(-h/2)$ = 0.4 nm, $\lambda_{S3}(h/2) = \lambda_{S3}(-h/2)$ = 40 nm.

### 4.2.3. Possible experimental signatures: pyroelectric coefficient profiles

The temperature dependence of improper and spontaneous polarization (via the temperature dependence of $A_i$ and $\beta_i^R(T, u_m, h)$) determines existence of pyroelectric response. **Figures 9** show the structural parameter, polarization and pyroelectric coefficient of short-circuited 20 nm SrTiO$_3$ film calculated for different temperatures and rather small positive misfit strain. It was shown earlier [11, 12], that improper polarization induced by the flexo-roto coupling in SrTiO$_3$ is temperature dependent and thus give rise to the surface or interface pyroelectric response. Here we demonstrate that the "improper" pyroelectric coefficient profile of thin SrTiO$_3$ films is anti-symmetric with respect to the film center. Theoretically estimated $\Pi_i$ values ~ $10^{-5} - 10^{-4}$ C/m$^2$K are comparable with the pyroelectric coefficients of known pyroactive materials like polyvinylidene difluoride (PVDF) and LiNbO$_3$ [41, 42].

In principal, detecting pyroelectric response of thin and ultra-thin film is possible by novel Pyro-SPM [43] performed at different temperatures. However, examination of pyroelectric activity thickness distribution by surface laser intensity modulation method (SLIMM) [42, 43] or photopyroelectric thermowave probing [44, 45] is problematic because the temperature wavelength $\lambda_T = \sqrt{a_T/\pi f}$ ($a_T$ is the thermal diffusivity, $f$ is the frequency of probing thermal flux modulation) should be less then the film thickness (~ 100 nm), that demands operation at rather high frequencies ~ 100 MHz. Nonzero pyroelectric response measured at wavelength $\lambda_T >> h$ can be registered under quasi-uniform temperature distribution entire the film thickness and gives information about the pyroelectric activity integrated over the film thickness. In the case of thin films (shown in **Figs.9a,b**), which are inversely polarized ($P_3(z) = -P_3(-z)$), the pyroelectric response measured in a sandwich electrode configuration should be negligibly small.

For thicker films when $\lambda_T < h/2$ at $f < 100$ MHz it is possible to register a pyroelectric response from near-surface unipolarized region connected with $P_3$ component (see **Figs. 9c,d**). Comparison of **Fig.9c** and **9d** shows that a strong pyroelectric asymmetry can be observed in thick films. At that, the asymmetry is not related with ***different*** extrapolation lengths (they are set equal for $z = \pm h/2$, i.e. $\lambda_{P1}(h/2) = \lambda_{P1}(-h/2)$, $\lambda_{P3}(h/2) = \lambda_{P3}(-h/2)$), but originated from misfit strain relaxation in thick enough films. For a film thickness more than several hundreds of nanometres misfit strain exists only in the vicinity of the clamped interface $z = +h/2$, and it vanishes when approaching mechanically free surface $z = -h/2$. The asymmetry in pyroactivity along film thickness allows registration of the total pyroelectric response by integral pyroelectric methods and pyroactivity probing by modulation methods.

The pyroelectric coefficients $\Pi_{1,2}$ caused by in-plane polarization components $P_{1,2}$ can be measured in planar electrode configurations [46], which give the information about pyroactivity arisen due to existence and interaction between the components $P_1$ and $P_2$. Such type of measurements were considered theoretically in [47] along with analysis of primary, secondary and tertiary pyroelectric effect contributions. Feng et al. [47] studied pyroelectric response of LiTaO$_3$ type II sensor with planar electrode configuration.

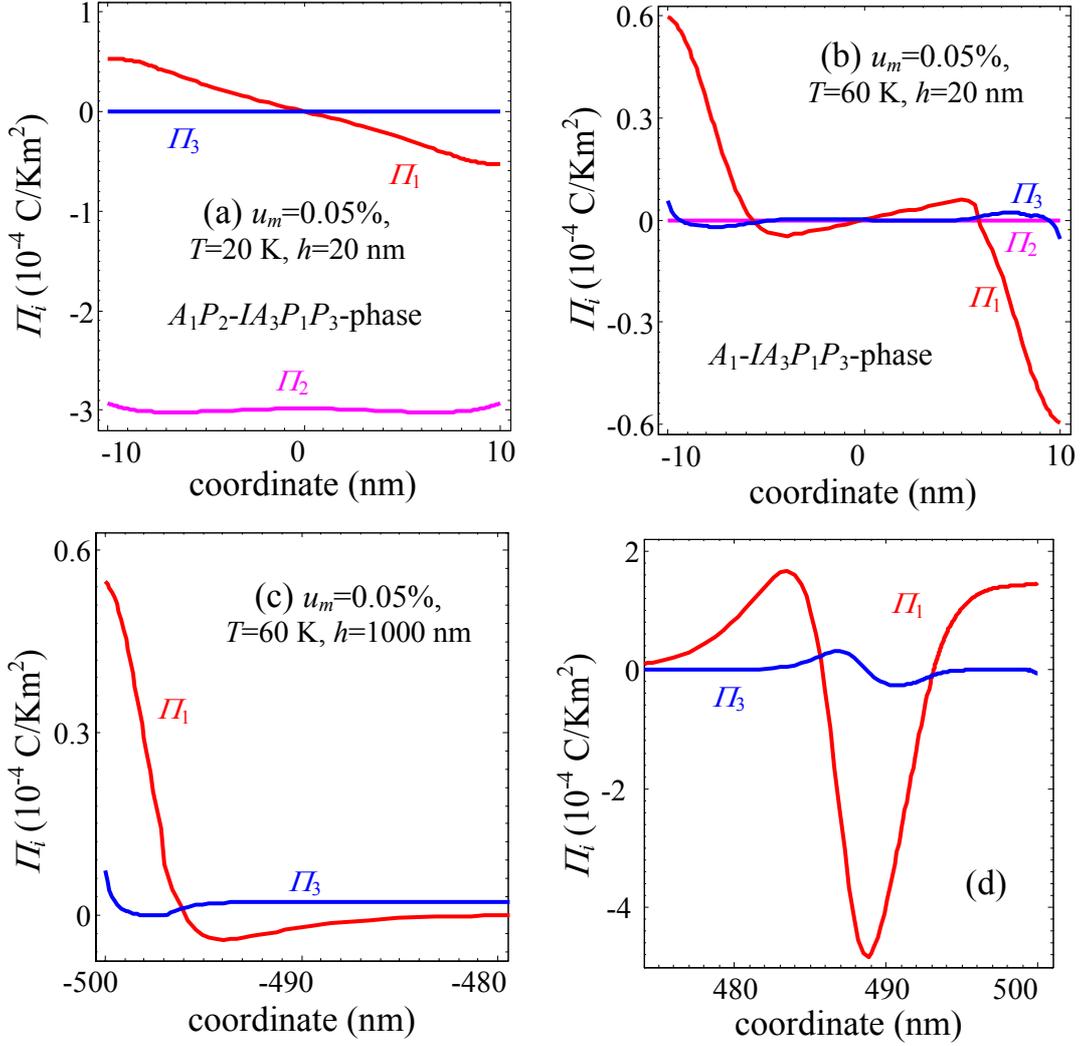

**Figure 9.** Distributions of pyroelectric coefficient components in a short-circuited $SrTiO_3$ film of thickness $h$=20 nm (a,b) and 1000 nm ((c) –near the interface $z = -h/2$, (d) –near the interface $z = +h/2$ ). Misfit strain $u_m$=0.05 % and temperature $T$=20 K (a, c, e) and 60 K (b, d, f). Extrapolation lengths are $\lambda_{P1}$=40 nm, $\lambda_{P3}$=40 nm, $\lambda_{S1}$=0.4 nm $\lambda_{S3}$=40 nm

## 5. Summary

Using Landau-Ginzburg-Devonshire (LGD) theory we study the influence of the flexoelectric and rotostriction coupling on the phase diagrams of ferroelastic-quantum paraelectric $SrTiO_3$ films. We determined phase diagrams in coordinates temperature - $SrTiO_3$ film thickness for different epitaxial misfit strains. Tensile misfit strains stimulate appearance of the spontaneous out-of-plane structural order parameter in the structural phase. Compressive misfit strains stimulate appearance of the spontaneous in-plane structural order parameter. Improper structural parameter components appear the vicinity of film surfaces as induced by the boundary conditions (surface energy).

Gradients of the structural order parameter components, which inevitably exist in the vicinity of film surfaces, can induce improper polarization and pyroelectric response via the flexoelectric and rotostriction coupling mechanism. Flexoelectric and rotostriction coupling results into the roto-flexoelectric field that is antisymmetric inside the film, small in the central part of the film, where the gradients of the structural parameter are small, and maximal near the surfaces, where the gradients of the structural parameter are the highest. The field induces antisymmetric improper polarization and pyroelectric response, which are zero in the central part of the film and maximal near the surfaces. Penetration depths of the improper phases (both polar and structural) can reach several nm from the film surfaces. Results obtained for $SrTiO_3$ films evidence in favor of the roto-flexo coupling importance in such oxide perovskites as $CaTiO_3$ and $EuTiO_3$ films.

The temperature dependence of improper and spontaneous polarization (via the temperature dependence of $A_i$ and $\beta_i^R(T, u_m, h)$) determines existence of pyroelectric response. Theoretically estimated $\Pi_i$ values ~ $10^{-5}$ - $10^{-4}$ C/m$^2$K are comparable with those of known pyroactive materials PVDF and $LiNbO_3$. The pyroelectric response originated from out-of-plane polarization component with symmetrical inversely polarized thickness distribution for thin films is negligible under thermal probing at frequencies $f$ <100 MHz. For thick enough films a strong pyroelectric asymmetry originated from misfit strain relaxation can be observed. Thus it becomes possible to register the total pyroelectric response by any conventional pyroelectric method, pyroactivity thickness distribution by photopyroelectric thermowave probing or SLIMM. The pyroelectric responses originated from the in-plane polarization components can be registered by conventional pyroelectric methods using planar electrode configurations.

**Acknowledgements**

To Prof. N.V. Morozovsky, Prof. A.K. Tagantsev and Prof. Venkatraman Gopalan for valuable suggestions and multiple discussions. Research supported (SVK, AB) by the U.S. Department of Energy, Basic Energy Sciences, Materials Sciences and Engineering Division.

**References**

[1]. J.H. Lee, Lei Fang, Eftihia Vlahos, Xianglin Ke, Young Woo Jung, L.F. Kourkoutis, Jong-Woo Kim, P.J. Ryan, Tassilo Heeg, M. Roeckerath, V. Goian, M. Bernhagen, R. Uecker, P.C. Hammel, K.M. Rabe, S. Kamba, J. Schubert, J.W. Freeland, D.A. Muller, C.J. Fennie, P. Schiffer, V. Gopalan, E. Johnston-Halperin, D. Schlom. Nature, **466**, 954, (2010) doi: 10.1038/nature09331

[2] Ji Young Jo, R. J. Sichel, E. M. Dufresne, Ho Nyung Lee, S. M. Nakhmanson, and P. G. Evans, Phys. Rev. B **82**, 174116 (2010)

[3] Ji Young Jo, R. J. Sichel, Ho Nyung Lee, S. M. Nakhmanson, E. M. Dufresne, and P. G. Evans, Phys. Rev. Lett. **104**, 207601 (2010)

[4] R. Nath, S. Zhong, S. P. Alpay, B. D. Huey, M. W. Cole, Appl. Phys. Lett. **92**, 012916 (2008)

[5] V. A. Stephanovich, I. A. Luk'yanchuk, and M. G. Karkut Phys. Rev. Lett. **94,** 047601 (2005)

[6] M.S. Majdoub, P. Sharma, and T. Cagin, Phys. Rev. **B 77**, 125424 (2008).

[7] E.A. Eliseev, A.N. Morozovska, M.D. Glinchuk, and R. Blinc. Phys. Rev. B. **79**, 165433, (2009).

[8] I. Ponomareva, A. K. Tagantsev, L. Bellaiche. Phys. Rev. B **85**, 104101 (2012).

[9] A.K. Tagantsev. Phys. Rev B, 34, 5883-5889 (1986).

[10] V. Gopalan and D.B. Litvin. Nature Materials **10**, 376-381 (2011).

[11] A.N. Morozovska, E.A. Eliseev, M.D. Glinchuk, Long-Qing Chen, Venkatraman Gopalan. Phys.Rev.B. **85**, 094107 (2012).

[12] A.N. Morozovska, E.A. Eliseev, S.V. Kalinin, Long-Qing Chen and Venkatraman Gopalan. Surface polar states and pyroelectricity in ferroelastics induced by flexo-roto field. *Accepted to APL* (http://arxiv.org/abs/1201.5085)

[13] N. A. Pertsev, A. K. Tagantsev, and N. Setter. Phys. Rev. B 61, R825 (2000).

[14] J. H. Haeni, P. Irvin, W. Chang, R. Uecker, P. Reiche, Y. L. Li, S. Choudhury, W. Tian, M. E. Hawley, B. Craigo, A. K. Tagantsev, X.Q. Pan, S.K. Streiffer, L. Q. Chen, S. W. Kirchoefer, J. Levy *&* D. G. Schlom. Nature **430**, 758 (2004).

15 Y. L. Li, S. Choudhury, J. H. Haeni, M. D. Biegalski, A. Vasudevarao, A. Sharan, H. Z. Ma, J. Levy, Venkatraman Gopalan, S. Trolier-McKinstry, D. G. Schlom, Q. X. Jia, and L. Q. Chen. Phys. Rev. B **73**, 184112 (2006).

16 G. Sheng, Y. L. Li, J. X. Zhang, S. Choudhury, Q. X. Jia, V. Gopalan, D. G. Schlom, Z. K. Liu, and L. Q. Chen. Appl. Phys. Lett. **96**, 232902 (2010).

[17] R. Kretschmer and K.Binder. Phys. Rev. B **20**, 1065 (1979).

[18] Chun-Lin Jia, Valanoor Nagarajan, Jia-Qing He, Lothar Houben, Tong Zhao, Ramamoorthy Ramesh, Knut Urban & Rainer Waser, Nature Materials, **6**, 64 (2007).

[19] Jiawang Hong and Daining Fang. Appl.Phys.Lett. 92,012906 (2008)

[20] Yihui Zhang, Jiawang Hong, Bin Liu, and Daining Fang, J. Appl. Phys. 108, 124109 (2010)

[21] P. Zubko, G. Catalan, A. Buckley, P.R. L. Welche, J. F. Scott. Phys. Rev. Lett. **99**, 167601 (2007).

[22] M. J. Haun, E. Furman, T. R. Halemane and L. E. Cross, Ferroelectrics, **99**, 55 (1989), ibidem p.13.

[23] . B. Houchmanzadeh, J. Lajzerowicz and E Salje, J. Phys.: Condens. Matter **3**, 5163 (1991)

[24] H. Uwe and T. Sakudo, Phys. Rev. B **13**, 271 (1976).

[25] J. H. Barrett, Phys. Rev. **86**, 118 (1952)

[26] A.K. Tagantsev, and G. Gerra, J. Appl. Phys. **100**, 051607 (2006).

[27] C.H. Woo and Yue Zheng, Appl. Phys. A **91**, 59 (2007)

[28] G. Rupprecht and R.O. Bell, Phys. Rev. **135**, A748 (1964).

[29] A. K. Tagantsev, A. S. Yurkov, Flexoelectric effect in finite samples, arXiv:1110.0380v1 [cond-mat.mtrl-sci] 3 Oct 2011

[30] A.S. Yurkov, JETP Letters, 94, No. 6, pp. 455–458 (2011). [Pis'ma v Zhurnal Eksperimental'noi i Teoreticheskoi Fiziki, 94, No. 6, pp. 490–493 (2011)]

[31] R. K. Behera, B. B. Hinojosa, S. B. Sinnott, A. Asthagiri, and S. R. Phillpot, J. Phys.: Condens. Matter, **20**, 395004 (2008).

[32] M. Sepliarsky, M. G. Stachiotti, and R. L. Migoni, Phys. Rev. **B, 72**, 014110 (2005)

[33] Yihui Zhang, Jiawang Hong, Bin Liu and Daining Fang, Nanotechnology 20, 405703 (2009)

[34] . Jiawang Hong, G. Catalan, D. N. Fang, Emilio Artacho, and J. F. Scott, Phys. Rev. B **81**, 172101 (2010).

[35] L.D. Landau, E.M. Lifshitz, and L.P. Pitaevskii, *Electrodynamics of Continuous Media*, 2nd ed. (Butterworth-Heinenann, Oxford, 1984) Vol. 8.

[36] A.K. Tagantsev, L.E. Cross, and J. Fousek: "Domains in Ferroic Crystals and Thin Films" Springer, 2010; Section 2.3 Thermodynamic approach, pgs. 81-84; and section 8.2 Basics of Domain State Reorientation, pgs. 354-360.

[37] See supplementary material at [URL will be inserted by AIP] for the calculations details.

[38] I.S. Gradshteyn and I.M. Ryzhik. Table of Integrals, Series, and Products, 5th ed edited by A Jeffrey. Academic, New York (1994)

[39] E.A. Eliseev, A.N. Morozovska. *The Journal of Materials Science*. 44, № 19, 5149-5160 (2009).

[40] W. Cao and R. Barsch, Phys. Rev. B, **41**, 4334 (1990).

[41] S.B. Lang. Sourcebook of Pyroelectricity. London - New York – Paris: Gordon & Breach Sci. Publishers (1974).

[42] S.L. Bravina, A.N. Morozovska, N.V. Morozovsky, S. Gille, J-P. Salvestrini, M. Fontana. Ferroelectrics, 353, 202-211 (2007).

[43] J. Groten, M. Zirkl, G. Jakopic, A. Leitner, and B. Stadlober, Phys. Rev. B **82**, 054112 (2010).

[44] S.L. Bravina, N.V. Morozovsky, A. A.Strokach, "Pyroelectricity: some new research and application aspects", in book *Material Science and Material Properties for Infrared Optoelectronics,* v. 3182, Ed. by F.F. Sizov, SPIE, Bellingham, pp.85-99 (1996).

[45] S. L. Bravina, N. V. Morozovsky, "Complete diagnostics of pyroactive structures for smart systems of optoelectronics", in book *Material Science and Material Properties for Infrared Optoelectronics,* v. 3359, Ed. By F.F. Sizov, SPIE, Bellingham, pp.420-427 (1998).

[46] A. Hadni. Infrared Physics 27, 17-23, (1987)

[47] Chuansheng Feng, Pingmao Xu. Infrared Physics and Technology 40, 61-78 (1999)

# Supplementary Materials to
# Roto-flexoelectric coupling impact on the phase diagrams and pyroelectricity of thin SrTiO$_3$ films

A.N. Morozovska[1,2], E.A. Eliseev[3], S.L. Bravina[1], A.Y. Borisevich[4], and S.V. Kalinin[5]

[1] *Institute of Physics, National Academy of Science of Ukraine,*
*46, pr. Nauki, 03028 Kiev, Ukraine*

[2] *Institute of Semiconductor Physics, National Academy of Science of Ukraine,*
*45, pr. Nauki, 03028 Kiev, Ukraine*

[3] *Institute for Problems of Materials Science, National Academy of Science of Ukraine,*
*3, Krjijanovskogo, 03142 Kiev, Ukraine*

[4] *Materials Sciences and Technology Division, Oak Ridge National Laboratory, Oak Ridge, TN 37831, USA*

[5] *Center for Nanophase Materials Science, Oak Ridge National Laboratory,*
*Oak Ridge, TN, 37831*

## A.1. Depolarization field

In the dielectric limit depolarization field $E_i^d$ obeys electrostatic equation:

$$\varepsilon_0 \varepsilon_b \frac{\partial E_3^d}{\partial x_3} = -\frac{\partial P_3}{\partial x_3}, \quad (A.1)$$

With boundary conditions corresponding to the limiting cases

(a) open-circuited boundary conditions, when electric displacement is zero outside the film. For the case $E_3^d = -P_3(x_3)/\varepsilon_0 \varepsilon_b$;

(b) short-circuited boundary conditions, when electric potential is zero outside the film. For the case $E_3^d = -(P_3(x_3) - \overline{P_3})/\varepsilon_0 \varepsilon_b$, where $\overline{P_3} = \frac{1}{h}\int_{-h/2}^{+h/2} dx_3 P_3(x_3)$.

## A.2. *Evident form of the equations of state and boundary conditions:*

Equations of state for the structural order parameter can be written as

$$2\left(\alpha_1^* + \frac{f_{11}r_{12}}{c_{11}}\frac{\partial P_3}{\partial x_3}\right)A_1 + 4\alpha_{11}^* A_1^3 + 2\alpha_{12}^* A_1 A_2^2 + 2\alpha_{13}^* A_1 A_3^2 - v_{44}\frac{\partial^2 A_1}{\partial x_3^2}$$
$$+ 2\left(\xi_{11}^* P_1^2 + \xi_{31}^* P_3^2 + \xi_{12}^* P_2^2\right)A_1 + \xi_{44} P_1 P_2 A_2 + \xi_{44}^* P_1 P_3 A_3 + \frac{f_{44} r_{44}}{c_{44}} A_3 \frac{\partial P_1}{\partial x_3} = 0 \quad (A.2a)$$

$$2\left(\alpha_1^* + \frac{f_{11}r_{12}}{c_{11}}\frac{\partial P_3}{\partial x_3}\right)A_2 + 4\alpha_{11}^*A_2^3 + 2\alpha_{12}^*A_2A_1^2 + 2\alpha_{13}^*A_2A_3^2 - v_{44}\frac{\partial^2 A_2}{\partial x_3^2}$$
$$+ 2\left(\xi_{11}^*P_2^2 + \xi_{31}^*P_3^2 + \xi_{12}^*P_1^2\right)A_2 + \xi_{44}P_1P_2A_1 + \xi_{44}^*P_2P_3A_3 + \frac{f_{44}r_{44}}{c_{44}}A_3\frac{\partial P_2}{\partial x_3} = 0 \quad \text{(A.2b)}$$

$$2\left(\alpha_3^* + \frac{f_{11}r_{11}}{c_{11}}\frac{\partial P_3}{\partial x_3}\right)A_3 + 4\alpha_{33}^*A_3^3 + 2\alpha_{13}^*(A_1^2 + A_2^2)A_3 - v_{11}\frac{\partial^2 A_3}{\partial x_3^2}$$
$$+ 2\left(\xi_{33}^*P_3^2 + \xi_{13}^*(P_1^2 + P_2^2)\right)A_3 + \xi_{44}^*(P_1A_1 + P_2A_2)P_3 + \frac{f_{44}r_{44}}{c_{44}}\left(A_1\frac{\partial P_1}{\partial x_3} + A_2\frac{\partial P_2}{\partial x_3}\right) = 0 \quad \text{(A.2c)}$$

Equations of state for polarization can be written as

$$2\beta_1^*P_1 + 4\beta_{11}^*P_1^3 + 2\beta_{12}^*P_1P_2^2 + 2\beta_{13}^*P_1P_3^2 + 2\left(\xi_{11}^*A_1^2 + \xi_{12}^*A_2^2 + \xi_{13}^*A_3^2\right)P_1$$
$$+ \xi_{44}A_2A_1P_2 + \xi_{44}^*A_1A_3P_3 - g_\parallel\frac{\partial^2 P_1}{\partial x_3^2} + 2\frac{f_{11}q_{12}}{c_{11}}\frac{\partial P_3}{\partial x_3}P_1 - \frac{f_{44}q_{44}}{c_{44}}P_1\frac{\partial P_3}{\partial x_3} = \frac{f_{44}r_{44}}{c_{44}}\frac{\partial(A_3A_1)}{\partial x_3} \quad \text{(A.2d)}$$

$$2\beta_1^*P_2 + 4\beta_{11}^*P_2^3 + 2\beta_{12}^*P_2P_1^2 + 2\beta_{13}^*P_2P_3^2 + 2\left(\xi_{11}^*A_2^2 + \xi_{12}^*A_1^2 + \xi_{13}^*A_3^2\right)P_2$$
$$+ \xi_{44}A_2A_1P_1 + \xi_{44}^*A_2A_3P_3 - g_\parallel\frac{\partial^2 P_2}{\partial x_3^2} + 2\frac{f_{11}q_{12}}{c_{11}}\frac{\partial P_3}{\partial x_3}P_2 - \frac{f_{44}q_{44}}{c_{44}}P_2\frac{\partial P_3}{\partial x_3} = \frac{f_{44}r_{44}}{c_{44}}\frac{\partial(A_3A_2)}{\partial x_3} \quad \text{(A.2e)}$$

For open-circuited electrical boundary conditions:

$$\left(2\beta_3^* + \frac{1}{\varepsilon_0\varepsilon_b}\right)P_3 + 4\beta_{33}^*P_3^3 + 2\beta_{13}^*P_3(P_1^2 + P_2^2) + 2\left(\xi_{33}^*A_3^2 + \xi_{31}^*(A_1^2 + A_2^2)\right)P_3$$
$$+ \xi_{44}^*(P_1A_1 + P_2A_2)A_3 - g_\perp\frac{\partial^2 P_3}{\partial x_3^2} - \frac{f_{11}q_{12}}{c_{11}}\frac{\partial(P_1^2 + P_2^2)}{\partial x_3} + \frac{f_{44}q_{44}}{c_{44}}\left(P_1\frac{\partial P_1}{\partial x_3} + P_2\frac{\partial P_2}{\partial x_3}\right) \quad \text{(A.2f)}$$
$$= \frac{\partial}{\partial x_3}\left(\frac{f_{11}r_{12}}{c_{11}}(A_1^2 + A_2^2) + \frac{f_{11}r_{11}}{c_{11}}A_3^2\right)$$

For short-circuited electrical boundary conditions:

$$\left(2\beta_3^* + \frac{1}{\varepsilon_0\varepsilon_b}\right)P_3 + 4\beta_{33}^*P_3^3 + 2\beta_{13}^*P_3(P_1^2 + P_2^2) + 2\left(\xi_{33}^*A_3^2 + \xi_{31}^*(A_1^2 + A_2^2)\right)P_3$$
$$+ \xi_{44}^*(P_1A_1 + P_2A_2)A_3 - g_\perp\frac{\partial^2 P_3}{\partial x_3^2} - \frac{f_{11}q_{12}}{c_{11}}\frac{\partial(P_1^2 + P_2^2)}{\partial x_3} + \frac{f_{44}q_{44}}{c_{44}}\left(P_1\frac{\partial P_1}{\partial x_3} + P_2\frac{\partial P_2}{\partial x_3}\right) \quad \text{(A.2g)}$$
$$= \frac{\partial}{\partial x_3}\left(\frac{f_{11}r_{12}}{c_{11}}(A_1^2 + A_2^2) + \frac{f_{11}r_{11}}{c_{11}}A_3^2\right) + \frac{\overline{P_3}}{\varepsilon_0\varepsilon_b}$$

Evident form of the boundary conditions for polarization:

$$\left(2b_1^sP_1 + n_3g_\parallel\frac{\partial P_1}{\partial x_3} + n_3\frac{f_{44}}{c_{44}}(q_{44}P_1P_3 + r_{44}A_1A_3)\right)\bigg|_S = 0, \quad \text{(A.3a)}$$

$$\left(2b_1^sP_2 + n_3g_\parallel\frac{\partial P_2}{\partial x_3} + n_3\frac{f_{44}}{c_{44}}(q_{44}P_2P_3 + r_{44}A_2A_3)\right)\bigg|_S = 0, \quad \text{(A.3b)}$$

$$\left( \begin{array}{l} 2b_3^S P_3 + n_3 g_\perp \dfrac{\partial P_3}{\partial z} + n_3 2\left( f_{12} - \dfrac{c_{12}}{c_{11}} f_{11} \right) u_m + \\ n_3 \dfrac{f_{11}}{c_{11}} \left( q_{11} P_3^2 + q_{12} P_2^2 + q_{12} P_1^2 + r_{11} A_3^2 + r_{12} A_2^2 + r_{12} A_1^2 \right) \end{array} \right)\Bigg|_S = 0. \qquad (A.3c)$$

Boundary conditions for the structural order parameter:

$$\left( 2a_1^S A_1 + n_3 v_{44} \dfrac{\partial A_1}{\partial x_3} \right)\Bigg|_S = 0, \quad \left( 2a_1^S A_2 + n_3 v_{44} \dfrac{\partial A_2}{\partial x_3} \right)\Bigg|_S = 0, \quad \left( 2a_3^S A_3 + n_3 v_{11} \dfrac{\partial A_3}{\partial x_3} \right)\Bigg|_S = 0 \qquad (A.4)$$

*A.3. Decoupling approximation*

Numerical analyses of the coupled system (A.2) showed that in the structural phase (e.g. octahedrally tilted phase), polarization value, orientation and corresponding derivatives weakly influence on the structural order parameters $A_i$ in such materials as perovskites $SrTiO_3$, $CaTiO_3$, $EuTiO_3$. For the case all polarization-dependent terms can be omitted in Eqs.(A.2a-c), and, in this *decoupling approximation,* they acquire the form:

$$2\alpha_1^* A_1 + 4\alpha_{11}^* A_1^3 + 2\alpha_{12}^* A_1 A_2^2 + 2\alpha_{13}^* A_1 A_3^2 - v_{44} \dfrac{\partial^2 A_1}{\partial x_3^2} = 0 \qquad (A.5a)$$

$$2\alpha_1^* A_2 + 4\alpha_{11}^* A_2^3 + 2\alpha_{12}^* A_2 A_1^2 + 2\alpha_{13}^* A_2 A_3^2 - v_{44} \dfrac{\partial^2 A_2}{\partial x_3^2} = 0 \qquad (A.5b)$$

Typically the gradient coefficient $v_{44} \gg v_{11}$ [ 1 ]. As a result, characteristic lengths $L_1^A(T, u_m) = L_2^A(T, u_m) = \sqrt{\dfrac{-v_{44}}{2\alpha_1^*(T, u_m)}}$, which determine the gradient scale of the structural order parameter components $A_1$ and $A_2$, are typically higher than lattice constant(s), but the length $L_3^A(T, u_m) = \left| \dfrac{v_{11}}{2\alpha_3^*(T, u_m)} \right|^{1/2}$ that determines the gradient scale of the structural order parameter component $A_3$ is (much) smaller than the lattice constant, thus

$$2\alpha_3^* A_3 + 4\alpha_{33}^* A_3^3 + 2\alpha_{13}^* \left( A_1^2 + A_2^2 \right) A_3 = v_{11} \dfrac{\partial^2 A_3}{\partial x_3^2} \approx 0 \qquad (A.5c)$$

Approximate analytical solution of Eqs.(A.5a,b) in the structural phase (i.e. at $\alpha_1^* < 0$) is:

$$A_1(x_3) = A_b \sqrt{\dfrac{2m}{1+m}} \operatorname{sn}\left( \dfrac{x_3}{L_1^A(T, u_m)\sqrt{1+m}} + c_0 \bigg| m \right), \quad A_2(x_3) = 0. \qquad (A.6a)$$

Here sn($u|m$) is the elliptical sine function. The amplitude $A_b = \sqrt{-\alpha_1^* / 2\alpha_{11}^*}$ and constants $m$ and $c_0$ should be determined from the symmetric boundary condition (A.4) at film surfaces $x_3 = \pm h/2$.

Symmetry of the conditions gives $c_0 = \dfrac{K(m)}{2}$, where K($m$) is complete elliptic integrals of the first kind [2]. Similar solution (1↔2) is

$$A_1(x_3) = 0, \quad A_2(x_3) A_b \sqrt{\dfrac{2m}{1+m}} \, \mathrm{sn}\left( \dfrac{x_3}{L_1^A(T,u_m)\sqrt{1+m}} + \dfrac{K(m)}{2} \middle| m \right). \tag{A.6b}$$

Solution, different from (A.6a-b) is

$$A_1(x_3) = A_2(x_3) = A_c \sqrt{\dfrac{2m}{1+m}} \, \mathrm{sn}\left( \dfrac{x_3}{L_1^A(T,u_m)\sqrt{1+m}} + \dfrac{K(m)}{2} \middle| m \right). \tag{A.6c}$$

Where the amplitude $A_c = \sqrt{-\alpha_1^*/(2\alpha_{11}^* + \alpha_{12}^* + \alpha_{13}^*)}$.

Approximate analytical solution of Eqs.(A.5c) is:

$$A_3(x_3) \approx \begin{cases} \sqrt{-\dfrac{\alpha_{13}^*(A_1^2(x_3) + A_2^2(x_3)) + \alpha_3^*}{2\alpha_{33}^*}}, & \alpha_{13}^*(A_1^2 + A_2^2) + \alpha_3^* < 0 \\ 0, & \alpha_{13}^*(A_1^2 + A_2^2) + \alpha_3^* > 0 \end{cases} \tag{A.7}$$

Free energy (4) minimum determines which of the solutions (A.6) or (A.7) is more energetically preferable. Then one should substitute expressions for $A_i$ into Eqs.(A.2d-g) for polarization.

Relatively simple results for polarization can be derived **open-circuited electrical boundary conditions**, since $P_3(x_3)$ appeared negligibly small due to the strong depolarization field. Moreover, corresponding spatial scale, $L_3^P(T,u_m) = \left| \dfrac{g_\perp}{2\beta_3^*(T,u_m) + 1/\varepsilon_0\varepsilon_b} \right|^{1/2}$, appeared ~0.1 nm, i.e. significantly smaller than the lattice constant (~0.4 nm) of considered perovskites. Thus, putting $P_2(x_3) \approx 0$, we simplify Eqs.(A.2d-e) as:

$$2\beta_1^* P_1 + 4\beta_{11}^* P_1^3 + 2\beta_{12}^* P_1 P_2^2 + 2(\xi_{11}^* A_1^2 + \xi_{12}^* A_2^2 + \xi_{13}^* A_3^2) P_1 + \xi_{44}^* A_2 A_1 P_2 = g_\parallel \dfrac{\partial^2 P_1}{\partial x_3^2} + \dfrac{f_{44} r_{44}}{c_{44}} \dfrac{\partial(A_3 A_1)}{\partial x_3}$$

(A.8a)

$$2\beta_1^* P_2 + 4\beta_{11}^* P_2^3 + 2\beta_{12}^* P_2 P_1^2 + 2(\xi_{11}^* A_2^2 + \xi_{12}^* A_1^2 + \xi_{13}^* A_3^2) P_2 + \xi_{44}^* A_2 A_1 P_1 = g_\parallel \dfrac{\partial^2 P_2}{\partial x_3^2} + \dfrac{f_{44} r_{44}}{c_{44}} \dfrac{\partial(A_3 A_2)}{\partial x_3}$$

(A.8b)

For the considered case either $A_1(x_3) = 0$, or $A_2(x_3) = 0$ or $A_1(x_3) = A_2(x_3) \neq 0$, and so Eqs.(A.8) allow evident simplifications. Characteristic lengths $L_1^P(T,u_m) = L_2^P(T,u_m) = \left| \dfrac{g_\parallel}{2\beta_1^*(T,u_m)} \right|^{1/2}$, which determine the gradient scale of the polarization components $P_1$ and $P_2$, are typically higher than

lattice constant(s). The fact allows us to solve Eqs.(A.8) along with substitution either (A.6) or (A.7) and boundary conditions (A.3a,b)

$$\left(2b_1^S P_1 \pm g_\| \frac{\partial P_1}{\partial x_3} \pm \frac{f_{44}}{c_{44}} r_{44} A_1 A_3 \right)\bigg|_{x_3=\pm h/2} = 0, \quad (A.9a)$$

$$\left(2b_1^S P_2 \pm g_\| \frac{\partial P_2}{\partial x_3} \pm \frac{f_{44}}{c_{44}} r_{44} A_2 A_3 \right)\bigg|_{x_3=\pm h/2} = 0, \quad (A.9b)$$

Analyses of equations (A.8) and conditions (A.9) along with expressions (A.6)-(A.7) lead to the conclusion that several polar phases are possible.

1) For the case $A_1(x_3) = A_2(x_3) \equiv A(x_3)$ and $P_1(x_3) = P_2(x_3) \equiv P(x_3)$, polarization $P(x_3)$ obeys equation:

$$\left(2\beta_1^* + \xi_{44} A^2 + 2\left((\xi_{11}^* + \xi_{12}^*)A^2 + \xi_{13}^* A_3^2\right)\right)P + \left(4\beta_{11}^* + 2\beta_{12}^*\right)P^3 - g_\| \frac{\partial^2 P}{\partial x_3^2} = \frac{f_{44} r_{44}}{c_{44}} \frac{\partial(A_3 A)}{\partial x_3}, \quad (A.10)$$

where $A_3(x_3) \approx \sqrt{-\frac{2\alpha_{13}^* A^2(x_3) + \alpha_3^*}{2\alpha_{33}^*}}$ for $2\alpha_{13}^* A^2 + \alpha_3^* < 0$, and $A_3(x_3) \approx 0$ for $2\alpha_{13}^* A^2 + \alpha_3^* < 0$.

Boundary condition is $\left(2b_1^S P \pm g_\| \frac{\partial P}{\partial x_3} \pm \frac{f_{44}}{c_{44}} r_{44} A A_3 \right)\bigg|_{x_3=\pm h/2} = 0$. Solution of Eq.(A.10) can be found numerically, but approximate analytics is possible for determination of the ferroelectric and polar phase boundaries. Ferroelectric phase boundary can be determined as the instability threshold of the homogeneous solution. Analytical expression for the homogeneous solution of Eq.(A.10) was obtained in the VKB-type approximation

$$P_{\text{hom}}(x_3) \approx C_1 \exp\left(i\int_0^{x_3}\sqrt{\frac{\gamma(x)}{-g_\|}}dx\right) + C_2 \exp\left(-i\int_0^{x_3}\sqrt{\frac{\gamma(x)}{-g_\|}}dx\right) \equiv C\cos\left(\int_0^{x_3}\sqrt{\frac{\gamma(x)}{-g_\|}}dx\right) \quad (A.11)$$

Extrapolation length $\lambda_P = \frac{g_\|}{2b_1^S}$, gradient coefficient $g_\| = g_{44} - \frac{f_{44}^2}{c_{44}}$, the even function $\gamma(x) = \left(2\beta_1^* + \xi_{44} A^2 + 2\left((\xi_{11}^* + \xi_{12}^*)A^2 + \xi_{13}^* A_3^2\right)\right)$. Substitution of Eq.(A.12) into the homogeneous boundary conditions $\left(P_1 \pm \lambda_P \frac{\partial P_1}{\partial x_3}\right)\bigg|_{x_3=\pm h/2} = 0$ gives the instability threshold as the condition of zero determinant

$$\cos\left(\int_0^{h/2}\sqrt{\frac{\gamma(x)}{-g_\|}}dx\right) - \lambda_P \sqrt{\frac{\gamma(h/2)}{-g_\|}} \sin\left(\int_0^{h/2}\sqrt{\frac{\gamma(x)}{-g_\|}}dx\right) = 0 \quad (A.12)$$

Transcendental Eq.(A.12) determines the ferroelectric phase boundary in coordinates film thickness $h$, misfit strain $u_m$ and temperature $T$. The boundary depends on the extrapolation length $\lambda_P$.

The boundary of the polar phase induced by the flexo-roto fields (inhomogeneities) is diffuse and can be estimated from the condition $\frac{\partial(A_3 A)}{\partial x_3} \approx 0$.

2) For the case $A_2(x_3) = 0$ and $A_1(x_3) \neq 0$ (or $A_1(x_3) = 0$ and $A_2(x_3) \neq 0$), $P_1(x_3) \neq 0$ and $P_2(x_3) = 0$ (or $P_1(x_3) = 0$ and $P_2(x_3) \neq 0$), we obtained that $P_1(x_3)$ obeys equation:

$$2\left(\beta_1^* + \xi_{11}^* A_1^2 + \xi_{13}^* A_3^2\right) P_1 + 4\beta_{11}^* P_1^3 - g_\parallel \frac{\partial^2 P_1}{\partial x_3^2} = \frac{f_{44} r_{44}}{c_{44}} \frac{\partial(A_3 A_1)}{\partial x_3}, \quad (A.13)$$

where $A_3(x_3) \approx \sqrt{-\frac{\alpha_{13}^* A_1^2(x_3) + \alpha_3^*}{2\alpha_{33}^*}}$ for $2\alpha_{13}^* A_1^2 + \alpha_3^* < 0$, and $A_3(x_3) \approx 0$ for $2\alpha_{13}^* A_1^2 + \alpha_3^* < 0$.

Boundary condition is $\left(2b_1^S P_1 \pm g_\parallel \frac{\partial P_1}{\partial x_3} \pm \frac{f_{44}}{c_{44}} r_{44} A_1 A_3\right)\bigg|_{x_3 = \pm h/2} = 0$. The ferroelectric phase boundary in coordinates film thickness $h$, misfit strain $u_m$ and temperature $T$ can be determined from Eq.(A.12), where $\gamma(x) = 2\left(\beta_1^* + \xi_{11}^* A_1^2(x) + \xi_{13}^* A_3^2(x)\right)$.

For the case of ***short-circuited electric boundary conditions*** the critical thickness of the out-of-plane ferroelectric polarization appearance can be determined from equation:

$$\cosh\left(\int_0^{h/2} \sqrt{\frac{\gamma(x)}{g_\perp}} dx\right) + \lambda_P \sqrt{\frac{\gamma(h/2)}{g_\perp}} \sinh\left(\int_0^{h/2} \sqrt{\frac{\gamma(x)}{g_\perp}} dx\right) = 0, \quad (A.14)$$

where $\gamma(x) = 2\beta_3^* + 2\left(\xi_{33}^* A_3^2 + \xi_{31}^* (A_1^2 + A_2^2)\right) + \frac{1}{\varepsilon_0 \varepsilon_b}$ and typically $\gamma(x) \approx \frac{1}{\varepsilon_0 \varepsilon_b}$.

## *A.4. Material parameters used in calculations*

**Table 1**. SrTiO$_3$ material parameters [3, 4].

| Parameter | SI units | Value | Source and notes |
|---|---|---|---|
| $\varepsilon_b$ | dimensionless | 43 | [a, b] |
| $\alpha_T$ | $10^6 \times$m/(F K) | 0.75 | [c, d] |
| $T_0^{(E)}$ | K | 30 | ibidem |
| $T_q^{(E)}$ | K | 54 | ibidem |
| $a_{ij}$ | $10^9 \times$m$^5$/(C$^2$F) | $a_{11}^u$=2.025, $a_{12}^u$=1.215, $a_{11}^\sigma$=1.724, $a_{12}^\sigma$=1.396 | ibidem calculated from $a_{ij}^u$ |
| $q_{ij}$ | $10^{10} \times$ m/F | $q_{11}$=1.251, $q_{12}$= −0.108, $q_{44}$=0.243 | [c] |
| $g_{ijkl}$ | $10^{-11} \times$V·m$^3$/C | $g_{11}$=$g_{44}$=1, $g_{12}$=0.5 | Estimation based on Ref. [e] |
| $\beta_T$ | $10^{26} \times$J/(m$^5$ K) | 9.1 | [c] |
| $T_S$ | K | 105 | [c] |
| $T_q^{(\Phi)}$ | K | 145 | [c] |
| $b_{ij}$ | $10^{50} \times$J/m$^7$ | $b_{11}^u$=1.94, $b_{12}^u$=3.96, $b_{11}^\sigma$=1.69, $b_{12}^\sigma$=3.88 | [c] calculated from $b_{ij}^u$ |

| $r_{ij}$ | $10^{30} \times$ J/(m$^5$) | $r_{11}$=1.3, $r_{12}$= –2.5, $r_{44}$= –2.3 | [c] |
|---|---|---|---|
| $\eta_{ijkl}$ | $10^{29}$ (F m)$^{-1}$ | $\eta_{11}^u$ = –3.366, $\eta_{12}^u$ = 0.135, $\eta_{44}^u$ =6.3 | [c] |
|  |  | $\eta_{11}^\sigma$ = –2.095, $\eta_{12}^\sigma$ = –0.849, $\eta_{44}^\sigma$ =5.860 | calculated from $\eta_{ij}^u$ |
| $v_{ijkl}$ | $10^{10} \times$ J/m$^3$ | $v_{11}$=0.28, $v_{12}$= –7.34, $v_{44}$=7.11 | [d, f] |
| $c_{ij}$ | $10^{11} \times$ J/m$^3$ | $c_{11}$=3.36, $c_{12}$=1.07, $c_{44}$=1.27 | [c, d] |
| $s_{ij}$ | $10^{-12} \times$ m$^3$/J | $s_{11}$=3.52, $s_{12}$= –0.85, $s_{44}$=7.87 | calculated from $c_{ij}$ |
| $f_{ijkl}$ | V | $f_{11}^e$ = – 3.24 , $f_{12}^e$ = 1.44 , $f_{44}^e$ = 1.08 | Recalculated from Ref.[g] at given stress |

[a] G. Rupprecht and R.O. Bell, Phys. Rev. **135**, A748 (1964).

[b] G.A. Smolenskii, V.A. Bokov, V.A. Isupov, N.N Krainik, R.E. Pasynkov, A.I. Sokolov, *Ferroelectrics and Related Materials* (Gordon and Breach, New York, 1984). P. 421

[c] N. A. Pertsev, A. K. Tagantsev, and N. Setter, Phys. Rev. B **61**, R825 (2000).

[d] A.K. Tagantsev, E. Courtens and L. Arzel, Phys. Rev. B, **64**, 224107 (2001).

[e] J. Hlinka and P. Marton, Phys. Rev. B **74**, 104104 (2006).

[f] W. Cao and R. Barsch, Phys. Rev. B, **41**, 4334 (1990)

[g] P. Zubko, G. Catalan, A. Buckley, P.R. L. Welche, J. F. Scott. Phys. Rev. Lett. **99**, 167601 (2007).

**References**

[1] W. Cao and R. Barsch, Phys. Rev. B, **41**, 4334 (1990).

[2] I.S. Gradshteyn and I.M. Ryzhik Table of Integrals, Series, and Products, 5th ed edited by A Jeffrey. Academic, New York (1994)

[3] A.N. Morozovska, E.A. Eliseev, M.D. Glinchuk, Long-Qing Chen, Venkatraman Gopalan. Phys.Rev.B. **85**, 094107 (2012)

[4] A.N. Morozovska, E.A. Eliseev, S.V. Kalinin, Long-Qing Chen and Venkatraman Gopalan. Surface polar states and pyroelectricity in ferroelastics induced by flexo-roto field. *Accepted to APL* (http://arxiv.org/abs/1201.5085)